\documentclass[11pt,letterpaper]{article}

\usepackage[style=authoryear-icomp, maxbibnames=9, maxcitenames=2]{biblatex}
\usepackage[T1]{fontenc}
\usepackage[utf8]{inputenc}
\usepackage{setspace}

\usepackage{microtype}

\usepackage{amsfonts,amsmath,amssymb,amsthm}

\usepackage{fontawesome}

\usepackage{booktabs}

\usepackage{subcaption}

\usepackage{wrapfig}

\usepackage{multicol}

\usepackage{pifont}
\newcommand{\cmark}{\ding{51}}%
\newcommand{\xmark}{\ding{55}}%

\usepackage[svgnames]{xcolor}

\usepackage[font={small,color=black,onehalfspacing}, labelfont={bf}]{caption}

\usepackage{graphicx}
\usepackage[export]{adjustbox}

\usepackage{csquotes}

\usepackage{url}
\usepackage{mwe}

\usepackage{enumitem}
\setlist[itemize]{topsep=0pt}
\setlist[enumerate]{topsep=0pt}

\usepackage{titlesec}
\titlespacing*{\section}
{0pt}{1ex plus 1ex minus .2ex}{1ex plus .2ex}
\titlespacing*{\subsection}
{0pt}{1ex plus 1ex minus .2ex}{1ex plus .2ex}

\usepackage{palatino}

\usepackage{lipsum}

\usepackage{floatrow}
\newfloatcommand{capbtabbox}{table}[][\FBwidth]

\newenvironment{sciabstract}{%
\begin{quote}\small\onehalfspacing}
{\end{quote}}

\title{
\vspace{-0mm}
Influence via Ethos: On the Persuasive Power of\\Reputation in Deliberation Online}

\author
{Emaad Manzoor,~
George H.~Chen,~
Dokyun Lee,~
Michael D. Smith\\
\normalsize{\{emaad,~georgechen,~dokyun,~mds\}@cmu.edu}\\
\normalsize{Carnegie Mellon University}\\[24pt]
}

\date{}

\begin{document}

\maketitle
\thispagestyle{empty}

\onehalfspacing

\begin{center}
\textbf{Abstract}
\end{center}

\begin{sciabstract}
  Deliberation among individuals online plays a key role
  in shaping the opinions that drive votes, purchases, donations and other critical offline behavior.
  Yet, the determinants of opinion-change via persuasion in deliberation online remain largely unexplored.
  Our research examines the persuasive power of \textit{ethos} -- an individual's ``reputation'' -- using a 7-year panel of over a million debates from an argumentation platform containing explicit indicators of successful persuasion.
  We identify the causal effect of reputation on persuasion by constructing an instrument
  for reputation from a measure of past debate competition, and by controlling
  for unstructured argument text using neural models of language in the double machine-learning framework.
  We find that an individual's reputation significantly impacts their persuasion rate above and beyond the
  validity, strength and presentation of their arguments.
  In our setting, we find that having 10 additional reputation points causes a 31\% increase in the probability
  of successful persuasion over the platform average.
  We also find that the impact of reputation is moderated by characteristics of the argument content,
  in a manner consistent with a theoretical model that attributes the persuasive
  power of reputation to heuristic information-processing under cognitive overload.
  We discuss managerial implications for platforms that facilitate deliberative decision-making for public and private organizations online.
\end{sciabstract}

\begin{quote}\small\textbf{Keywords:} Persuasion, reputation systems, double machine-learning, causal inference from text\end{quote}

\makeatletter{\renewcommand*{\@makefnmark}{}
\footnotetext{Preliminary draft, comments are welcome.}\makeatother}

\clearpage
\pagenumbering{arabic}

\section{Introduction}

Deliberation --- ``an extended conversation among two or more people in order to come to a better understanding of some issue'' \parencite{beauchampmodeling} -- forms the grease of societal decision-making machinery, lubricating consensus among participants via fair and informed debate. The process of opinion exchange in deliberation alleviates polarization, minority under-representation and several other drawbacks of consensus formation arising from non-deliberative processes (such as in majority-voting without discussion) by educating potentially uninformed participants and broadening their awareness of alternative perspectives \parencite{list2013deliberation,thompson2008deliberative}.

An increasing amount of deliberation takes place online \parencite{davies2009online}, both on social media and on specialized platforms developed for participatory democracy\footnote{For example, see the Stanford Online Deliberation Platform: \url{https://stanforddeliberate.org/}}, knowledge curation\footnote{For example, see Wikipedia Talk pages used to discuss Wikipedia edits: \url{https://en.wikipedia.org/wiki/Help:Talk_pages}} and software planning\footnote{For example, see Github Issues used to plan open-source projects: \url{https://guides.github.com/features/issues/}}, among others. While going virtual broadens participation, the increased visibility of ``reputation'' indicators online could distort the equitability of the deliberation process. For example, \parencite{marlow2013impression} find that project managers on Github (used for open-source software development by Google, Facebook and Microsoft, among several other technology firms) use visible reputation indicators when evaluating users' feature requests and critiquing developers' code contributions. At the same time, this creates opportunities for firms to exploit reputation indicators when directly interacting with consumers online to promote sales and mitigate churn (as gaming giant Electronic Arts does on Reddit, for example). These opportunities also extend to philanthropic organizations engaged in curbing the spread of misinformation online.

\begin{figure*}[!h]
  \vspace{2.5mm}
  \centering
  \includegraphics[width=\linewidth]{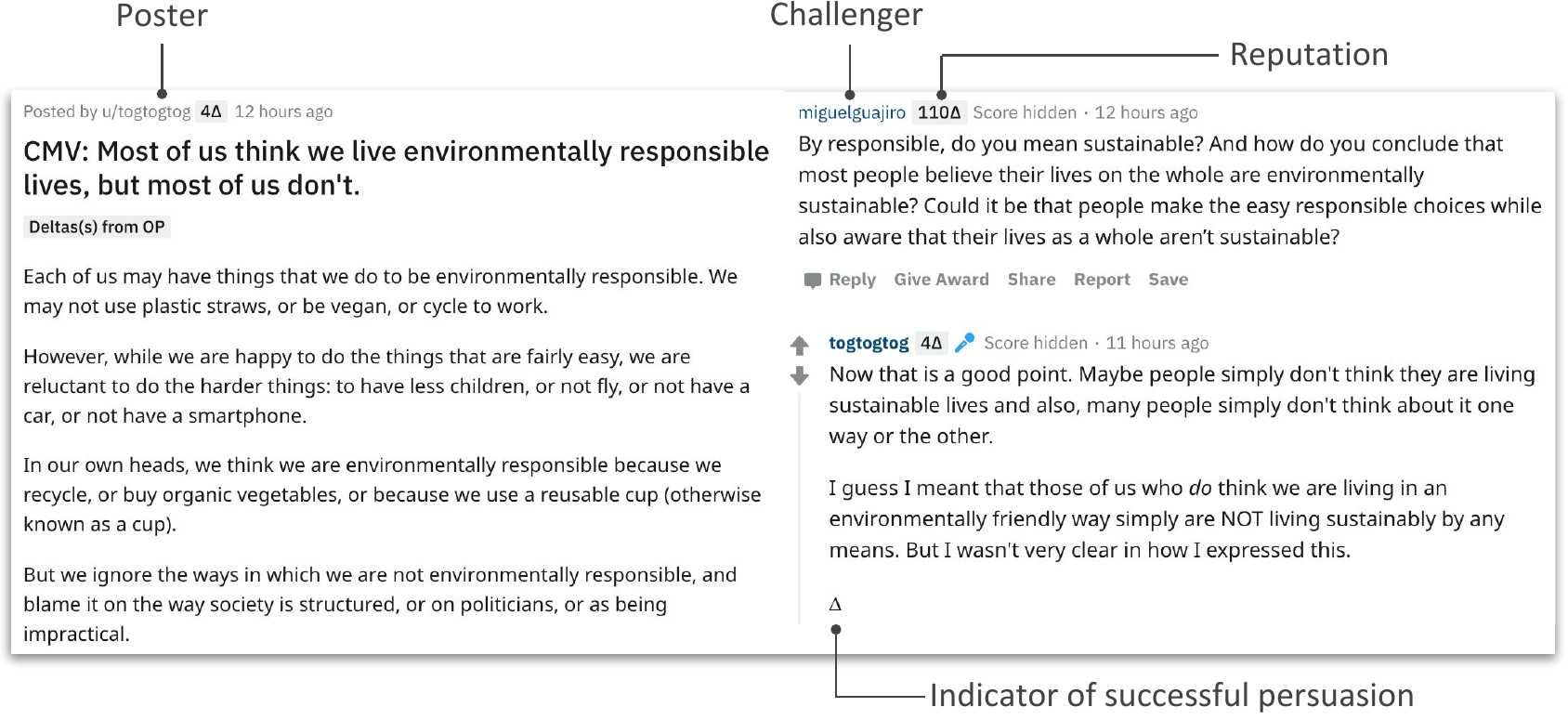}
  \caption{\textbf{A debate.} An opinion shared by poster \textit{togtogtog} (left), a response by challenger \textit{miguelguajiro} (top-right) and a reply by \textit{togtogtog} indicating successful persuasion with the $\Delta$ symbol (bottom-right). Displayed above \textit{miguelguajiro}'s response is their reputation (110$\Delta$), which is the number $\Delta$s earned previously.
  \vspace{5mm}}
  \label{fig:cmv-screenshot}
\end{figure*}

Whether reputation indeed has persuasive power in online deliberation is thus an important concern, but one that is difficult to quantify due to the difficulty of recognizing opinion-change and persuasion even on the rare occasions when it does occur. We overcome this challenge by assembling a dataset of deliberation from the ChangeMyView\footnote{http://reddit.com/r/changemyview/} online argumentation platform, containing over a million debates spanning 7 years from 2013 to 2019. Strict curation by a team of over 20 moderators ensures that debates on ChangeMyView are well-informed, balanced and civil, thus satisfying the key tenets of authentic deliberation \parencite{fishkin2005experimenting}. The debates in our dataset cover a variety of topics, from politics and religion to comparisons of products and brands, reflecting the diverse interests of over 800,000 ChangeMyView users.
ChangeMyView users initiate debates by sharing opinions, engage in dyadic deliberation with other users that challenge their opinion, and (uniquely) provide \textit{explicit indicators of successful persuasion} for each challenger that persuaded them to change their opinion. For every user persuaded, challengers earn reputation points that are prominently displayed with their username on the platform. The screenshot in Figure \ref{fig:cmv-screenshot} illustrates the deliberation process and the nature of reputation indicators on ChangeMyView.

\clearpage

\noindent
We use this dataset to analyze whether an individual's reputation impacts their persuasiveness in deliberation online, beyond the
content of their arguments.
Our identification strategy to answer this question draws
on four key components, enabled by several unique characteristics of our dataset:
\begin{enumerate}[topsep=6pt, itemsep=6pt, label=\textbf{\Roman*}.]
  \item \textbf{Within-opinion variation}: We exploit the availability of multiple challengers of each opinion to analyze within-opinion variation (via opinion fixed-effects). This controls for unobserved characteristics of the opinion (such as the topic) and the poster (such as their agreeability) that may introduce biases arising from users endogenously selecting which opinions to challenge.

  \item \textbf{Approximating persuasive ability:} We exploit the availability of multiple persuasion attempts for each user over time to measure and control for their past (lagged) persuasion rate, as a proxy for their unobserved persuasive ability (or skill) in each debate.

  \item \textbf{Instrumenting for reputation:} We derive an instrument for the reputation of the challenger in each debate to address potential confounding due to unobserved challenger characteristics that vary over time, and are hence not controlled for by their past persuasion rate.

  \item \textbf{Controlling for the response text:} Each challenger's response text is the primary medium through which their persuasive ability, linguistic fluency and other major determinants of persuasion are observed by the poster. By controlling for the response text nonparametrically, we control for and address potential confounding arising from all such determinants.
\end{enumerate}

\clearpage

We instrument for the challenger's reputation in each debate with their average position in the sequence of responses to opinions they challenged previously (their \textit{mean past position}). For a given opinion, challengers responding earlier (at lower positions) exhaust the limited space of good arguments, making it harder for challengers responding later (at higher positions) to persuade the poster\footnote{This resembles the mechanism of the cable news channel position instrument used to quantify the persuasive power of Fox News on voting Republican \parencite{martin2017bias}.}. Hence, we expect challengers with higher (worse) mean past positions to have lower reputations in the present, motivating our instrument's relevance. While users can strategically select opinions to challenge that have fewer earlier challengers, our instrument remains exogenous after controlling for the user's present position in each debate. To further alleviate concerns of instrument validity, we derive conservative bounds on our estimates with relaxed instrument validity assumptions using the plausibly-exogenous instrumental variable framework \parencite{conley2012plausibly}.

Text plays a key role in ensuring instrument validity. All confounders of the instrument must affect both the instrument and the debate outcome (whether the poster was persuaded). To affect the debate outcome, such confounders must operate through channels observable by the poster, the most prominent of which is the text of the challenger's response. Hence, ``controlling for'' the challenger's response text blocks the causal pathways between such confounders and the debate outcome, ensuring that they do not violate instrument validity.

To operationalize this intuition in an ideal world, we would manually annotate, measure and control for every possible characteristic of the response text that could affect the debate outcome, which is infeasible at scale. An alternative is to control for a bag-of-words\footnote{A bag-of-words representation of a document is a high-dimensional vector of the frequencies of all the words it contains. Its dimensionality is the size of the vocabulary of words in the document corpus, which is typically of the order of millions.} \parencite{harris1954distributional} vector of the response text, assuming that functions of this vector capture all text characteristics that determine the debate outcome. However, the high dimensionality of bag-of-words representations introduces statistical difficulties that prevent consistent estimation and valid inference.

Dimensionality-reduction techniques are commonly employed to alleviate these difficulties, whether manually via hand-selected features, or automatically via inverse-regression \parencite{taddy2013multinomial}, topic-modeling \parencite{blei2003latent,roberts2016model,roberts2018adjusting} and neural text embeddings \parencite{mikolov2013distributed}. However, these techniques provide no guarantees that the confounders present in the original text are retained in the low-dimensional text representation, which raises concerns of omitted variable bias. In addition, there is often little substantive theory to guide the manual feature selection process. Automated dimensionality reduction techniques, including supervised ones such as feature selection via LASSO \parencite{tibshirani1996regression}, could result in inconsistent estimates due to model misspecification and invalid confidence intervals due to feature selection uncertainty that is not accounted for in the inference procedure \parencite{belloni2014high}.

\clearpage

We depart from the focus on dimensionality-reduction and instead incorporate the response text as a control nonparametrically, using recent advances in semiparametric inference with machine learning models. Specifically, we estimate ``nuisance functions'' of the response text via machine learning to predict the debate outcome, challenger reputation and instrument, and partial-out their effects in the manner of Frisch-Waugh-Lovell \parencite{frisch1933partial,lovell1963seasonal}. This procedure was introduced as early as \parencite{robinson1988root} for parametric nuisance functions and recently extended to nonparametric nuisance functions estimated via machine learning \parencite{van2003unified,chernozhukov2018double}. The recent extensions show that the partialling-out procedure guarantees $\sqrt{n}$-consistent and asymptotically normal estimates, as long as each estimated nuisance function converges to the true nuisance function at the rate of $n^{-1/4}$ or better.

In particular, we use a recent econometric extension of the partialling-out procedure called double machine-learning \parencite{chernozhukov2018double} to estimate a partially-linear instrumental variable specification with text as a control. For our nuisance functions, we use neural networks with rectified linear unit (ReLU) activation functions \parencite{nair2010rectified}. These neural networks pass the input text through a series of intermediate layers, each of which learns a latent ``representation'' that captures textual semantics at different granularities. The networks are trained via backpropogation \parencite{rumelhart1986learning} with first-order gradient-based techniques \parencite{kingma2015adam} to minimize classification or regressions loss functions. Though recurrent \parencite{hochreiter1997long} and convolutional \parencite{kim2014convolutional} neural networks are more commonly used for textual prediction tasks, neural networks with ReLU activation functions come with guaranteed $n^{-1/4}$ convergence rates \parencite{farrell2018deep} that enable consistent estimation and valid inference in the double machine-learning framework.

\noindent \textbf{Results.} We find a significant positive effect of reputation on persuasion. Our instrumental variable estimates indicate that having 10 additional units of reputation increases the probability of persuading a poster by 1.09 percentage points. This corresponds to a 31\% increase over the platform average persuasion rate of 3.5\%. Since each poster successfully persuaded increases a challenger's reputation, the long-run effect of reputation on persuasion is compounded over time. The effect remains statistically significant across a range of specifications, including ones where the instrument exclusion restriction is relaxed. Our findings counter the prevailing notion on the ChangeMyView platform that the persuasive power of reputation can be ignored.

The estimated effect of reputation on persuasion is the local average treatment effect (LATE) \parencite{imbens1994identification} in the population of \textit{compliers}, comprised of debates where the challenger's reputation (the treatment) is affected by their mean past position (the instrument). Such challengers are less persuasive at higher (later, worse) response positions and more persuasive at lower (earlier, better) response positions. Hence, we expect debates in the complier population to involve challengers with moderate to high persuasive ability, since challengers with low persuasive ability are unlikely to be any more persuasive at any response position.

To investigate possible mechanisms for this effect, we test the predictions of a theoretical model of persuasion with information-processing shortcuts called reference cues \parencite{bilancini2018rational}. We examine how the proportional effects of a challenger's reputation and skill vary with characteristics of the opinion and response content. Using the challenger's response text length as a proxy for the cognitive complexity of their arguments, we find that the reputation effect share (of the total effect magnitude of reputation and skill) increases from 82\% to 89\% from the first to the fourth response length quantile. This suggests that posters rely more on reputation when the challenger's arguments are cognitively complex. This is consistent with the theoretical prediction that individuals will rely more on low-effort heuristic processing (using reputation as a proxy for the quality of the challenger's response) instead of high-effort systematic processing (directly evaluating the challenger's response) when subject to greater cognitive overload.

The theoretical model also predicts that individuals will rely less on low-effort heuristic processing when they are more involved in the issue being debated. We test this prediction using the opinion text length as a proxy for the issue-involvement of the poster and find that the reputation effect share decreases from 90\% to 83\% from the second to the fourth opinion length quantile. This is consistent with the prediction that more issue-involved posters will rely less on reputation. We find similar patterns using text complexity measures (such as the Flesch-Kincaid Reading Ease) as proxies for cognitive complexity and issue-involvement, instead of the response and opinion text length. Overall, our findings are consistent with reputation serving as a reference cue and used by posters as an information-processing shortcut under cognitive overload.

We also examine how the effect of reputation on persuasion is moderated by the total number of opinion challengers. While we expect that having more challengers will increase the cognitive burden placed on the poster (and hence push them to rely more on heuristic information-processing), we find no evidence that posters rely more on reputation as the number of opinion challengers increases. We do find evidence that challengers with higher reputation have longer conversations with posters, which could be an important mediator of the effect of reputation on persuasion. We also find evidence that challengers with higher reputation are more likely to attract collaboration from other (non-poster) users, although reputation continues to have a significant (positive) direct effect on persuasion after excluding the potential effect of such collaboration.

\noindent \textbf{Contributions and related work.} Our research contributes to the economics of persuasion, the practitioners of which comprise over a quarter of the United States' GDP \parencite{mccloskey1995one}, including lawyers, judges, lobbyists, religious workers and salespeople. \parencite{antioch2013persuasion} revises this number to 30 percent, after including marketing, advertising and political campaigning professionals. An extensive body of past work on persuasion spans the economics, marketing and political science literature (among others), and is comprised of both theoretical models \parencite{kamenica2011bayesian,kamenica2018bayesian}
and empirical analyses of the efficacy of persuasive communication via field or natural experiments (see \cite{dellavigna2010persuasion} for a survey).

Our work differs from previous empirical studies on the economics of persuasion in three ways. First, previous work focused on identifying the existence of persuasion by quantifying the causal effect of persuasive communication on some observable behavior, without the ability to observe individual-level opinion-change. Content and persuader-based moderators of persuasion were then analyzed conditional on non-zero persuasive effects having been identified \parencite{landry2006toward,bertrand2010s}. In our work, the explicit indicators of persuasion provided by posters allows us to sidestep the task of identifying persuasion, and directly analyze its determinants.
Second, we observe attempts at persuasion made by thousands of unique individuals, in contrast with previous work. This enables a broader investigation of the impact of persuader and content characteristics, which are predicted to play an important role by belief-based persuasion models \parencite{stigler1961economics,mullainathan2008coarse,kamenica2011bayesian}. Finally, we observe repeated attempts at persuasion made by each individual that enables approximating and disentangling the impact of their persuasive ability from other factors.

More specifically, our work informs persuasive information design \parencite{kamenica2018bayesian} in interactive settings by quantifying the impact of extraneous signals that could serve as low-effort information-processing heuristics \parencite{petty1986elaboration,chaiken1989heuristic,todorov2002heuristic}.
Such heuristics play an increasingly important role in this era of information overload \parencite{jones2004information}, as emphasized by Cialdini in his seminal book on the principles of influence \parencite{cialdini2007influence}:
\begin{displayquote}
  \vspace{-1mm}
  "Finally, each principle is examined as to its ability to produce a distinct kind of automatic, mindless compliance from people, that is, a willingness to say yes without thinking first. The evidence suggests that the ever-accelerating pace and informational crush of modern life will make this particular form of unthinking compliance more and more prevalent in the future. It will be increasingly important for the society, therefore, to understand the how and why of automatic influence."
  \vspace{-1mm}
\end{displayquote}

\indent Interactive persuasion channels are common today, with firms adopting online channels such as live-chat to triangulate consumers' beliefs and influence them via dialogue. Interactive channels are often preferred for defensive marketing tasks \parencite{hauser1983defensive} such as addressing complaints and mitigating churn. Some firms invest in interaction further and embed themselves as bonafide members of influential enthusiast-run online forums\footnote{A notable examples is gaming giant Electronic Arts (\url{https://www.reddit.com/user/EACommunityTeam/}).}.
Marketing communication designed to persuade in such channels closely resembles the dyadic deliberation we examine in our work.

Our work is also related to research on the impact of certification and reputation systems \parencite{dranove2010quality} in markets for labor \parencite{moreno2014doing,kokkodis2016reputation}, knowledge \parencite{dev2019quantifying}, and other goods and services \parencite{tadelis2016reputation,hui2016reputation,lu2018can}. Consumers studied in this line of research engage in costly information-processing to evaluate item quality under cognitive, temporal or financial constraints.
Hence, the findings therein are interpreted using the same underlying psychological mechanisms as we employ in our work \parencite{petty1986elaboration,chaiken1989heuristic}. The distinguishing feature of our work is the focus on explicitly stated opinion-change as the outcome, as a consequence of interpersonal deliberation. Importantly, there are no monetary transactions involved and the reputation in our setting cannot be purchased at any cost; it is a truthful proxy for past persuasive ability. Thus, persuasion as exhibited in our setting is sufficiently different from the purchasing or hiring decisions analyzed in the literature on certification and reputation systems to warrant separate investigation.

Our work complements studies on deliberation in online settings, such as on political forums and social media \parencite{beauchampmodeling,shugars2019keep}. Specifically, our findings contribute to the understanding of opinion-change and polarization\footnote{https://www.wsj.com/articles/to-get-along-better-we-need-better-arguments-1531411024}
online \parencite{quattrociocchi2016echo}. By quantifying how an individual's reliance on heuristic and systematic information-processing varies with the cognitive complexity of the persuasive message content, our findings could inform online campaigns that involve persuasive information design aimed at reducing polarization by affecting opinion-change.

Finally, our work contributes an application to the nascent study of causal inference from text, and more broadly to the literature on text as data \parencite{gentzkow2019text,netzer2019words,toubia2019extracting}. Our setting involves text as a control (see \cite{keith2020text} for a recent survey of work in this setting).
Previous approaches to accommodate text as a control (though with treatments assumed to be exogenous) include \parencite{sridhar2019estimating} which controls for topics in the text, \parencite{roberts2018adjusting} which assumes a structural topic model \parencite{roberts2016model} of text and controls for its sufficient reduction \parencite{taddy2013multinomial}, and \parencite{veitch2019using,shi2019adapting} which incorporate neural language models of text in the targeted learning inference framework \parencite{van2011targeted}.

Our work also links the social science literature on persuasion with the computational natural language processing literature on argument-mining \parencite{lippi2016argumentation}, where online argumentation platforms have been extensively studied \parencite{tan2016winning,jo2018attentive,luu2019measuring,atkinson2019gets,srinivasan2019content}.

\noindent \textbf{Outline.} We begin in Section \ref{sec:conceptual} by introducing background, formalizing our conceptual framework and motivating our hypotheses. We then describe our dataset in Section \ref{sec:data} and detail our empirical strategy in Section \ref{sec:empirical}, including a description of our estimation procedure and evidence supporting the validity of our instrument. We discuss our results in Section \ref{sec:results} and interpret them through the lens of a theoretical model of persuasion. We conclude by summarizing our findings, discussing managerial implications for platforms facilitating online deliberation for public and private organizations, and noting the limitations of our research in Section \ref{sec:conclusion}.

\clearpage

\section{Background and Conceptual Framework}
\label{sec:conceptual}

The ChangeMyView online argumentation platform was created in January, 2013 to foster good-faith discussions on polarizing issues and has received praise for helping combat the proliferation of echo chambers online\footnote{``Civil discourse exists in this small corner of the internet'' --- The Atlantic. December 30, 2018.}. In this section, we  formalize the process of deliberation on ChangeMyView and describe important platform features to motivate our empirical analyses in Section \ref{sec:empirical}.

\noindent \textbf{Opinion posters, opinion challengers and debates.} Our unit of analysis is a \textit{debate}. Each debate is associated with an opinion shared by an opinion \textit{poster}, which is titled with the poster's primary claim and contains at least 500 characters of supporting arguments. A response to the opinion by a \textit{challenger} initiates a debate between the poster and challenger. Other users can (but rarely) join the ongoing discussion between a poster and a challenger with their own comments; we term such debates \textit{multi-party}. Debates must follow several rules (detailed in Appendix A) enforced by over 20 moderators. Notable rules are: (i) the poster must personally hold a non-neutral opinion, (ii) the poster must engage with all challengers for at least 3 hours after sharing their opinion, and (iii) a challenger's response must counter at least one claim made by the poster. Responses to an opinion are ordered chronologically and popularity votes on responses are hidden for the first 24 hours after an opinion is shared. These rules mitigate popularity biases, irrelevant digressions and hostility.

\noindent \textbf{Opinion selection by users.} The titles of posted opinions and the identities of the posters who shared them are displayed in a paginated list on the platform's homepage, ordered by a combination of recency and popularity votes\footnote{Specifically, in decreasing order of the score: $\textrm{sign}(\textit{upvotes} - \textit{downvotes})\textrm{log}_{10}|\textit{upvotes} - \textit{downvotes}| + \textit{post-datetime}/45000$.}. A tab on the homepage also allows users to order opinions by recency only. Clicking on an opinion title opens a new page displaying the opinion text and any ongoing or concluded debates between the poster and other challengers. Users could select opinions to challenge based on various factors such as the opinion text, their own topical preferences, the poster's identity, and the number and status of the debates between the poster and other challengers.

\noindent \textbf{The $\Delta$-system.} In mid-February, 2013, ChangeMyView introduced a reputation system called the $\Delta$-system to incentivize challenging opinions on the platform. At any point in a debate, the poster may reply to the challenger indicating that their opinion has changed using the $\Delta$ symbol or equivalent alternatives. We term debates where the poster awarded a $\Delta$ to the challenger as \textit{successful} and opinions that led to at least one successful debate as \textit{conceded}. Due to the platform rules requiring active engagement, 98\% of the $\Delta$s from the poster in our dataset were awarded within 24 hours of the opinion being posted, with over 50\% being awarded within just 90 minutes. This short delay reduces concerns of opinion-change occurring due channels external to the debate. Each awarded $\Delta$ grants the challenger a reputation point. Other non-poster users can (but rarely) also award $\Delta$s to any challenger and contribute to their reputation. The total reputation points earned previously, if non-zero, are displayed next to the challenger's username with all of their responses on the platform.

\noindent \textbf{The poster's decision.} Consider an opinion $p$ that is challenged by user $u$. The poster observes $u$'s username, reputation $r_{pu}$ and the text of their immediate response to the opinion. Based on this information, the poster may initiate a discussion with the challenger, elicit additional responses (which we do not model) and eventually award a $\Delta$ if persuaded to change their opinion. We model the poster $p$'s decision to award a $\Delta$ to challenger $u$ as a function of an opinion-specific threshold $\tau_p$ and the perceived quality $\tilde{q}_{pu}$ of $u$'s response:
\begin{align}
  Y_{pu} ~~ &= ~~ \mathbb{I}[\tilde{q}_{pu} > \tau_p]       & \mathbb{I}[x] = 1 \textrm{ if } x \textrm{ is true }, \mathbb{I}[x] = 0 \textrm{ otherwise}\nonumber\\
  \tilde{q}_{pu} ~~ &= ~~ \alpha_r r_{pu} + \alpha_q q_{pu} & \alpha_r + \alpha_q = 1
  \label{eq:ohdecision}
\end{align}
\indent
Here, $Y_{pu} = 1$ is the observed debate outcome if the poster awarded a $\Delta$ to $u$ and $Y_{pu} = 0$ otherwise. The unobserved threshold $\tau_p$ encodes opinion-specific characteristics such as the opinion topic and the poster's openness to persuasion.
Based on \parencite{dewatripont2005modes,bilancini2018rational}, we model the perceived quality $\tilde{q}_{pu}$ as a weighted linear combination of the challenger's reputation $r_{pu}$ and the ``true'' response quality $q_{pu}$, which the poster can determine by evaluating the challenger's response at some cognitive cost. Posters choose $\alpha_r$ and $\alpha_q$ endogenously based on this cognitive cost and their reliance on heuristic and systematic information-processing \parencite{petty1986elaboration,chaiken1989heuristic}.
If $\alpha_r > 0$, reputation in this model serves as a \textit{reference cue}: a proxy for the true response quality that can be processed with lesser effort than evaluating $q_{pu}$ directly.

\noindent \textbf{``True'' response quality.} We model the true response quality $q_{pu}$ as a function of the user's ``skill'' $s_{pu}$ at the time they challenged opinion $p$ and their position $t_{pu}$ in the sequence of challengers of opinion $p$. $t_{pu}$ captures the overall impact of previous challengers' responses. For example, challengers responding earlier could exhaust the limited space of good arguments, making it harder for later challengers to respond with arguments of similar quality. We formalize this as follows:
\begin{align}
  t_{pu} ~~ &= ~~\sum_{u'} \mathbb{I}[u' \textrm{ challenged opinion } p \textrm{ before $u$}]\nonumber\\
  q_{pu} ~~ &= ~~ \gamma_s s_{pu} + \gamma_t t_{pu}
  \label{eq:quality}
\end{align}
\noindent We approximate $u$'s skill by the Laplace-smoothed \parencite{manning2008introduction}
fraction of posters persuaded before opinion $p$, where $p$ is chronologically-ordered and $S_{p'u}=1$ if $u$ challenged opinion $p'$:
\begin{align}
  s_{pu} = \frac{
    \sum_{p' < p} Y_{p'u}S_{p'u} + 2s_{\mu}
  }{
    \sum_{p' < p} S_{p'u} + 2
  }
  \label{eq:skill}
\end{align}
Here, $s_{\mu}$ is a ``prior'' set to the empirical persuasion probability of users in their first debate ($\approx 1.6\%$). Smoothing ensures that the skill of users measured when they have challenged few opinions tends to $s_{\mu}$ instead of to 0.
A user's skill is thus their (smoothed) lagged persuasion rate, which captures all user characteristics that affect persuasion and do not change with the their tenure on the platform.

\noindent \textbf{Hypotheses.} Based on prior analytical work \parencite{bilancini2018rational}, we test three complementary hypotheses on the weights $\alpha_r$ and $\alpha_q$, which reflect the poster's endogenously-determined reliance on heuristic and systematic information-processing respectively:
\begin{enumerate}[label=\textit{H\arabic*.}, leftmargin=0.57in, rightmargin=1.5in, listparindent=\parindent, itemsep=12pt]
 \item \textit{Reputation has persuasive power}, $\alpha_r > 0$.
 \item \textit{The relative persuasive power of reputation, $\frac{\alpha_r}{\alpha_r + \alpha_q}$, increases
               as the cognitive cost of processing the challenger's response increases.}
 \item \textit{The relative persuasive power of reputation, $\frac{\alpha_r}{\alpha_r + \alpha_q}$, decreases
               as the involvement of the poster in the debated issue increases.}
\end{enumerate}
Confirming (H1) indicates that reputation has persuasive power, and confirming (H2) and (H3) lends support to the mechanism proposed by the model of (Bilancini and Boncinelli, 2018).
\section{Data}
\label{sec:data}

We collect all the discussions on the ChangeMyView platform between January, 2013 and October, 2019 using a combination of the official Reddit API\footnote{https://www.reddit.com/dev/api/} and the third-party PushShift API \parencite{baumgartner2020pushshift}, in full compliance with their terms of service. We exclude submissions to ChangeMyView that are not opinions using the fact that opinion titles are required to be prefixed with ``CMV:''. The excluded submissions encompass discussions about the platform, announcements of platform changes and celebrations of milestones. We also exclude the opinions and responses posted to ChangeMyView before the reputation system became fully functional on March 1, 2013.

\begin{wrapfigure}[15]{r}{0.3\textwidth}
  \vspace{-5mm}
  \includegraphics[width=\linewidth]{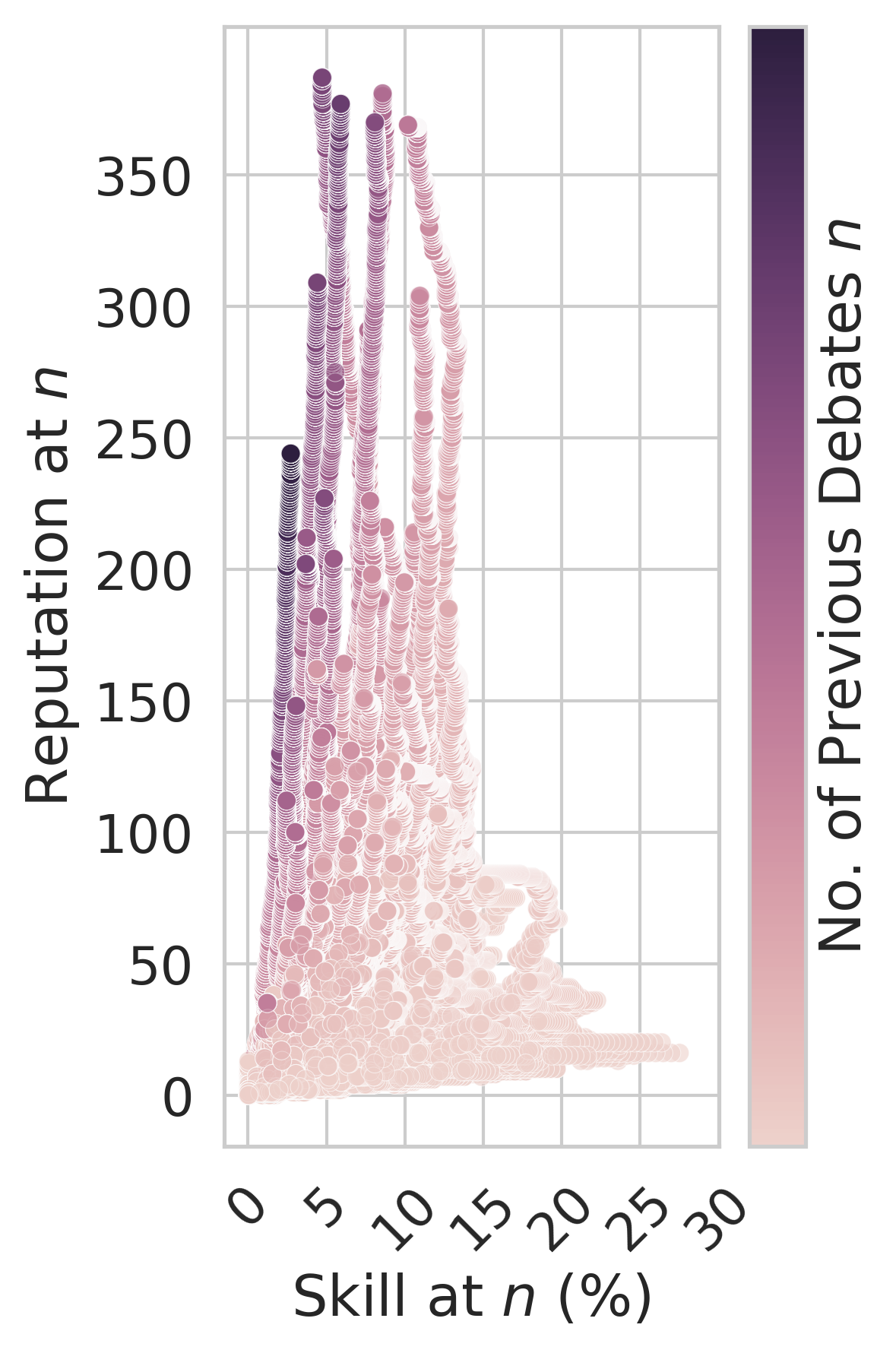}
  \caption{Reputation and skill}
  \label{fig:cmv-skill0}
\end{wrapfigure}
\noindent We extract indicators of successful persuasion from the debate text using the same extraction rules employed by ChangeMyView to programmatically parse $\Delta$s and other alternative symbols\footnote{Code obtained from: https://github.com/alexames/DeltaBot}. We use the extracted indicators to label debate success, to reconstruct each challenger's reputation and to measure each challenger's skill in each debate.
Figure \ref{fig:cmv-skill0} shows the empirical variation in skill with reputation in our dataset, with each point indicating the reputation and skill for each challenger measured in a single debate, colored based on the number of debates they participated in previously. At values of skill outside the low and high extremes, there is a wide variation in the reputation ($r=0.29, p<0.001$). This variation is essential to disentangle the effects of reputation and skill on persuasion.

\clearpage

\begin{figure*}[!h]
  \centering
  \begin{subfigure}[t]{0.45\linewidth}
    \includegraphics[width=\linewidth]{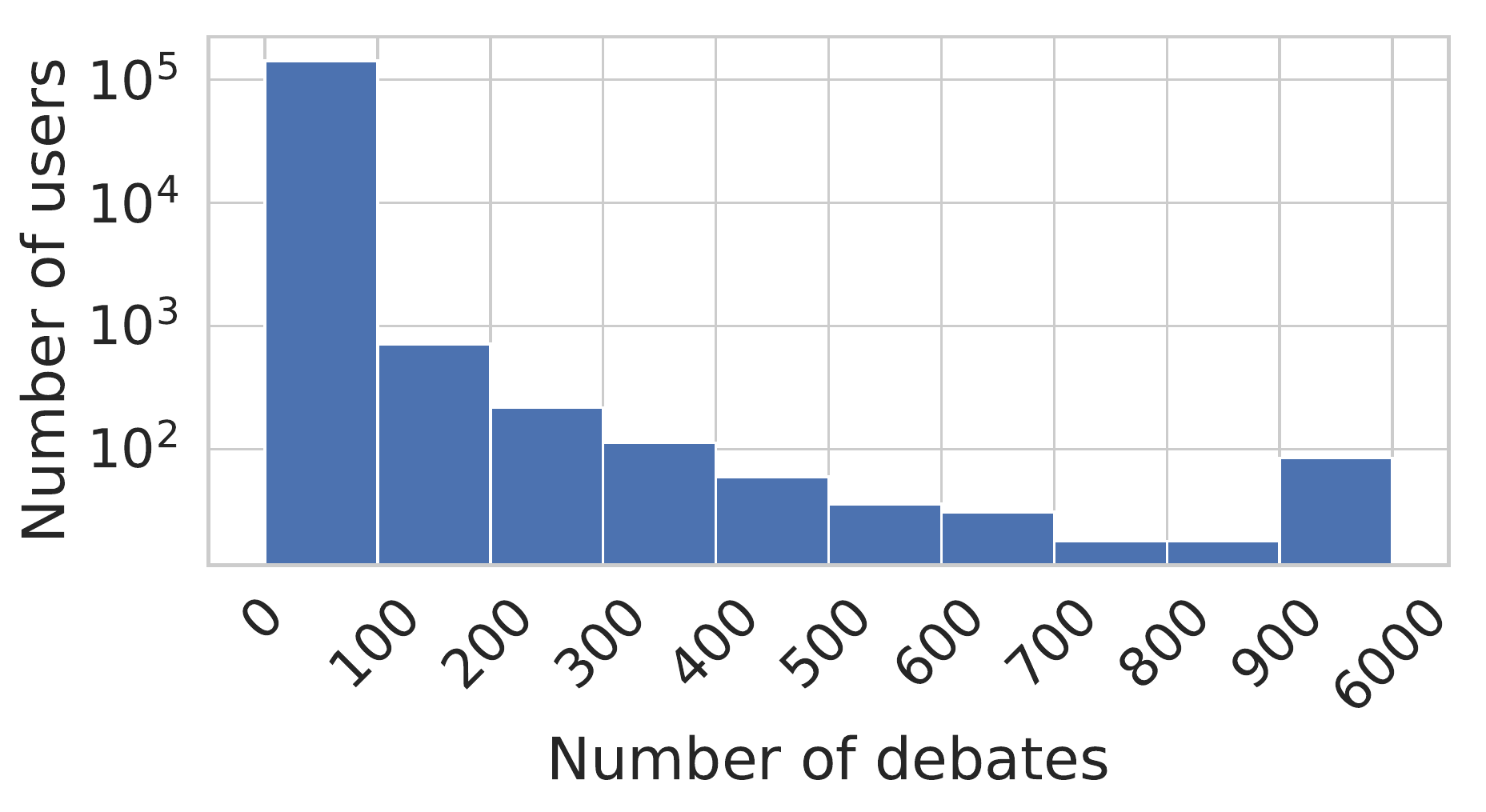}
    \label{fig:users-attempts}
  \end{subfigure}
  \begin{subfigure}[t]{0.45\linewidth}
    \includegraphics[width=\linewidth]{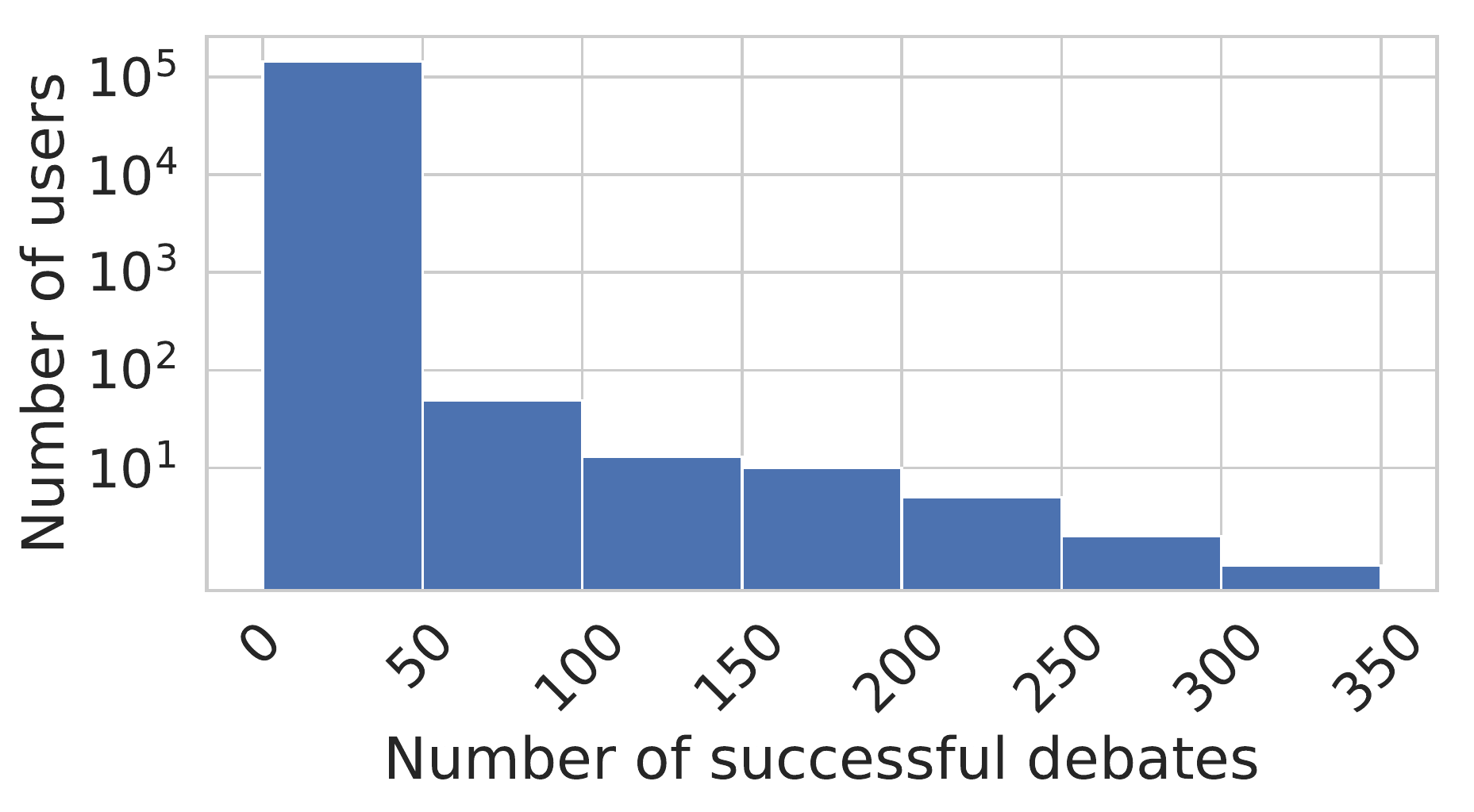}
    \label{fig:users-successes}
  \end{subfigure}
  \vspace{-7.5mm}
  \caption{\textbf{Debate participation and success.} Distribution of total and successful debates per user.}
  \label{fig:users}
\end{figure*}

Debates by challengers who had deleted their ChangeMyView accounts before data collection appear in our dataset with the ``[deleted]'' placeholder username. The inability to link the debates by such challengers over time makes it impossible to measure their true reputation and skill. Assuming that such challengers have zero reputation and skill $s_{\mu}$ (based on equation \ref{eq:skill}) is likely to attenuate our estimates due to measurement error. Hence, we exclude all 118,277 such debates from our dataset\footnote{For completeness, we also report our main results including debates with deleted challengers in Appendix C.}. Our final dataset contains 91,730 opinions (23.5\% of them conceded) shared by 60,573 unique posters, which led to 1,026,201 debates (3.5\% of them successful) with 143,891 unique challengers. Table \ref{tab:descriptive} reports descriptive statistics of our dataset, and Figure \ref{fig:users} reports user-level distributions of participation and debate success. Table \ref{tab:notation} summarizes the notation that will use in all subsequent sections.

\addtolength{\tabcolsep}{1mm}
\begin{table*}[!h]
\small
\centering
\caption{\textbf{Descriptive Statistics.} Debates from March 1, 2013 to October 10, 2019.}
\begin{tabular}[t]{lrrrr}
\toprule
         &Mean & Standard Deviation & Median \\
\midrule
\multicolumn{3}{l}{\textit{Statistics of challengers in each debate}}\\
Reputation $r_{pu}$                              & 15.9             & 43.4  & 1.0  \\
Skill $s_{pu}$ (\%)                              & 3.0              & 3.7   & 1.6  \\
Position $t_{pu}$                                & 14.8             & 24.3  & 8.0  \\
Mean past position $Z_{pu}$                      & 10.4             & 13.0  & 7.5  \\
Number of past debates $\sum_{p'<p} S_{p'u}$     & 244.4            & 591.7 & 24.00 \\
\midrule
\multicolumn{3}{l}{\textit{Statistics of overall dataset}}\\
Number of opinions                               &91,730            &        &      \\
\qquad Opinions conceded                         &21,576            &        &      \\
\qquad Opinions leading to more than 1 debate    &84,998            &\multicolumn{2}{l}{\textit{(number of clusters with opinion fixed-effects)}}\\
Number of debates                                &1,026,201         &        &      \\
\qquad Successful debates                        &36,187            &        &      \\
\qquad Multi-party debates                       &348,041           &        &      \\
Number of debates per opinion                    &11.2              & 12.7   & 9    \\
\qquad Successful debates per opinion            &0.4               & 0.9    & 0    \\
Number of unique posters                         &60,573            &        &      \\
\qquad Opinions per poster                       &1.5               & 2.4    & 1    \\
Number of unique challengers                     &143,891           &        &      \\
\qquad Challengers with more than 1 debate       &64,871            &\multicolumn{2}{l}{\textit{(number of clusters with user fixed-effects)}}\\
Number of debates per challenger                 &7.1               & 58.5   & 1    \\
\qquad Successful debates per challenger         &0.3               & 3.2    & 0    \\
\bottomrule
\end{tabular}
\label{tab:descriptive}
\end{table*}%
\addtolength{\tabcolsep}{-1mm}

\clearpage

{ %
\addtolength{\tabcolsep}{3mm}
\renewcommand{\arraystretch}{1.0}
\begin{table*}[t]
\vspace{-7.5mm}
\small
\centering
\caption{\textbf{Summary of notation.} List of recurring symbols introduced in Sections \ref{sec:conceptual} and \ref{sec:empirical}.}
\begin{tabular}[t]{ll}
\toprule
Symbol & \\
\midrule
$p$                & Chronological opinion index \\
$u$                & Chronological response index \\
$pu$               & Tuple representing a debate: the $u^{\textrm{th}}$ response to opinion $p$ \\
$\tau_p$           & Opinion fixed-effect; captures unobserved opinion characteristics \\
$\rho_u$           & Challenger fixed-effect; captures unobserved challenger characteristics \\
$r_{pu}$           & Reputation of the challenger in debate $pu$; sum of the past $\Delta$s earned\\
$s_{pu}$           & Skill of the challenger in debate $pu$; smoothed lagged persuasion rate \\
$t_{pu}$           & Position of the challenger in debate $pu$ \\
$m_{pu}$           & Calendar month-year fixed-effect for debate $pu$ \\
$X_{pu}$           & Vector representation of the text of the challenger's immediate response in debate $pu$  \\
$Y_{pu}$           & Binary outcome of debate $pu$; $Y_{pu}=1$ for successful debates, $Y_{pu}=0$ otherwise \\
$S_{pu}$           & Binary opinion selection indicator; $S_{pu}=1$ if $u$ challenged opinion $p$, $S_{pu}=0$ otherwise \\
$Z_{pu}$           & Instrument (mean past position) for the challenger's reputation in debate $pu$ \\
\bottomrule
\end{tabular}
\label{tab:notation}
\end{table*}%
\addtolength{\tabcolsep}{-3mm}
}
~\\[-20.5mm]
\section{Empirical Strategy}
\label{sec:empirical}
~\\[-20mm]
\subsection{Baseline Specifications}
\label{sec:baseline}
~\\[-9mm]
Equations (\ref{eq:ohdecision}) and (\ref{eq:quality}) motivate baseline specifications that relate the observed debate outcome $Y_{pu}$ to the challenger's reputation $r_{pu}$, skill $s_{pu}$ and position $t_{pu}$ as follows (constants omitted for brevity):
\begin{align*}
  Y^*_{pu} &= \tau_p + \beta_1 r_{pu} + \beta_2 s_{pu} + \beta_3 t_{pu} + \epsilon_{pu}
               &\mathbb{E}[\epsilon_{pu}|\tau_p, r_{pu}, s_{pu}, t_{pu}] = 0\\
  Y_{pu} &= \mathbb{I}[Y^*_{pu} > 0]
\end{align*}
\indent Here, $\tau_p$ is an opinion fixed-effect and $\epsilon_{pu}$ is an error term with zero conditional mean. Since the fixed-effects are at the opinion level and skill (a function of lagged dependent variables) is at the user level, these are not dynamic panel specifications, and are hence unaffected by Nickell bias \parencite{nickell1981biases}. Including the opinion fixed-effects excludes 6,732 debates from the sample, which were the only responses to their respective opinions.
If distributional assumptions (such as Gumbel or Gaussian) on $\epsilon_{pu}$ hold and there are no unobserved confounders, the estimate of $\beta_1$ quantifies the change in the probability of persuading the poster of opinion $p$ upon increasing the challenger's reputation by one unit, with all else equal. In Section \ref{sec:results}, we report estimates from logistic and linear probability models.

While the assumption of no unobserved confounding is restrictive (and relaxed in Section \ref{sec:instrument}), the baseline specifications address two important sources of confounding. First, controlling for the challenger's skill controls for all challenger characteristics that affect persuasion (such as their rhetorical ability and linguistic fluency) and that do not vary with their tenure on ChangeMyView. To see why such characteristics confound the effect of reputation on persuasion, note that a user's reputation largely depends on the number of posters persuaded previously: $r_{pu} \approx \sum_{p'<p} Y_{p'u}$ (since users who are not posters rarely award $\Delta$s). Hence, any unobserved challenger characteristic that affects the outcome of every debate $Y_{pu}, Y_{(p-1)u}, Y_{(p-2)u}, \dots$ will also affect their reputation $r_{pu} \approx Y_{(p-1)u} + Y_{(p-2)u} + \dots$, and thus confound the effect of reputation $r_{pu}$ on the debate outcome $Y_{pu}$.

\addtolength{\tabcolsep}{5mm}
\begin{table*}[!t]
\vspace{-1mm}
\small
\centering
\caption{\textbf{Estimated effect of past experience on debate success.}}
\begin{tabular}[t]{lc}
\toprule
& \multicolumn{1}{c}{Dependent Variable: Debate Success $Y_{pu}$}\\
\midrule
No. of opinions challenged previously $\sum_{p'<p}S_{p'u}$      & $-1 \times 10^{-6}$ ($0.7 \times 10^{-6}$)      \\
Position $t_{pu}$ (std. deviations)                             & $-0.0107$ ($0.0003$)$^{***}$                     \\
User fixed-effects ($\rho_u$)                                   & \cmark                                          \\
Month-year fixed-effects ($m_{pu}$)                             & \cmark                                          \\
No. of debates                                                  & $947,181$                                       \\
$R^2$                                                           & $0.07$                                          \\
\bottomrule
\end{tabular}
\label{tab:experience}
\vspace{3mm}

Note: Standard errors displayed in parentheses. $^{***}p<0.001; ^{**}p<0.01; ^{*}p<0.05$
\end{table*}%
\addtolength{\tabcolsep}{-5mm}

However, skill does not capture challenger characteristics that vary with their tenure on ChangeMyView. By assuming the absence of such characteristics, the baseline specifications implicitly assume that users do not learn to be more persuasive with experience on the platform. We provide empirical evidence to support this assumption by estimating the following linear probability model:
\begin{align*}
  Y_{pu} =  \rho_u + m_{pu}
          + \theta_1 \sum_{p' < p} S_{p'u} + \theta_2 t_{pu} + \epsilon_{pu}
\end{align*}
where $\rho_u$ is a user fixed-effect capturing all unobserved time-invariant user characteristics, $m_{pu}$ is a calendar month-year fixed-effect capturing unobserved temporal factors, $t_{pu}$ is the (standardized) user's position in the sequence of challengers of opinion $p$ and $\epsilon_{pu}$ is a Gaussian error term. $\sum_{p' < p} S_{p'u}$ is the number of opinions that $u$ challenged previously, serving as a measure of their past experience. $\theta_1$ is the within-user correlation between past experience and the debate outcome.
If users improve with experience, we expect $\theta_1$ to be positive. However, the estimates of $\theta_1$ reported in Table \ref{tab:experience} are small and statistically insignificant. We attribute this to users having already acquired argumentation experience outside the platform, with little to gain from additional experience on the platform.

Second, controlling for the opinion fixed-effect $\tau_p$ addresses confounding due to users endogenously selecting which opinions to challenge. To see why opinion selection is a concern, recall the opinion selection indicator $S_{pu}$ that equals 1 when user $u$ challenges opinion $p$. Since we estimate our specifications on observed debates, our specifications implicitly condition on $S_{pu}=1$. If the opinion selection probability $\mathbb{P}[S_{pu}=1]$ is correlated with (i) reputation, and (ii) debate success (for example, if users prefer to challenge opinions on topics that are easier to persuade in), the effect of reputation on debate success will be confounded due to endogenous sample selection \parencite{james1979sample}.

We characterize this confounding using the causal graph in Figure \ref{fig:sampleselection} based on the analyses in \parencite{hernan2004structural}. In causal graphs \parencite{pearl2009causality}, an edge $A\rightarrow B$ implies that $A$ \textit{may or may not cause} $B$, while the absence of an edge implies the stronger assumption that $A$ \textit{does not cause} $B$. An undirected edge $\leftrightarrow$ implies potential causality in either direction. Observed variables are shaded and unobserved variables are unshaded.

\begin{minipage}{\linewidth}
  \hspace{-18mm}
  \begin{minipage}[b]{0.522\linewidth}
    \centering
    \includegraphics[width=3in]{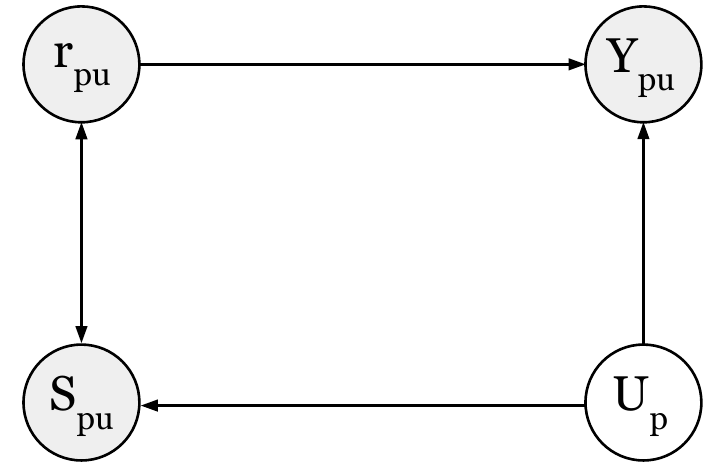}
    \captionof{figure}{\textbf{Opinion selection causal graph.}}
    \label{fig:sampleselection}
  \end{minipage}
  \begin{minipage}[b]{0.57\linewidth}
    \small
    \centering
    \begin{tabular}[t]{lc}
    \toprule
    \multicolumn{2}{c}{
      Dependent Variable: $\mathbb{I}[u \textrm{ challenges > 1 future opinion}]$
      }\\
    \midrule
    Reputation $r_{pu}$ (10 units)                & $0.0177$ ($0.0001$)$^{***}$ \\
    User fixed-effects ($\rho_u$)                 & \cmark \\
    Month-year fixed-effects ($m_{pu}$)           & \cmark \\
    No. of debates                                & $947,181$ \\
    $R^2$                                         & $0.56$    \\
    \bottomrule
    \end{tabular}
    \label{tab:sample-selection-treatment}
    \vspace{3mm}

    Note: Standard errors displayed in parentheses.

    $^{***}p<0.001; ^{**}p<0.01; ^{*}p<0.05$

    \captionof{table}{\textbf{Reputation/opinion selection correlation.}}
    \label{tab:participation}
    \end{minipage}
\end{minipage}

The shaded nodes $r_{pu}$, $Y_{pu}$ and $S_{pu}$ correspond to the reputation, debate outcome and opinion selection indicator respectively.
The unshaded node $U_{p}$ is any unobserved opinion characteristic that could directly affect both opinion selection and debate success, such as the opinion topic.

In the causal graph in Figure \ref{fig:sampleselection}, $S_{pu}$ is a \textit{collider}. A collider is any node $C$ that is a common outcome in causal substructures of the form $X \rightarrow C \leftarrow Y$. Conditioning on $C$ opens a causal pathway between $X$ and $Y$ that would otherwise be blocked.
If reputation is correlated with opinion selection (depicted by the undirected edge $r_{pu} \leftrightarrow S_{pu}$), conditioning on the collider $S_{pu}$ (which our specifications do implicitly) opens the confounding causal pathway $U_{p} \rightarrow S_{pu} \leftrightarrow r_{pu}$. This confounds the effect of reputation on the debate outcome, since $U_{p}$ now affects both $Y_{pu}$ and $r_{pu}$ (via $S_{pu}$).

We test for correlation between $r_{pu}$ and $S_{pu}$ by estimating the following linear probability model of a user challenging more than one opinion after opinion $p$:
\begin{align*}
  \mathbb{I}[u \textrm{ challenges > 1 future opinion}] = \rho_u + m_{pu} + \theta_1 r_{pu} + \epsilon_{pu}
\end{align*}
where $\rho_u$ is a user fixed-effect, $m_{pu}$ is a calendar month-year fixed-effect and $\epsilon_{pu}$ is a Gaussian error term.
The estimate of $\theta_1$ in Table \ref{tab:participation} suggests a significant positive correlation between $r_{pu}$ and $S_{pu}$. This correlation may arise either because users that were successful in the past (and hence have higher reputation) are more likely to challenge opinions in the future, or because more active users are likely to have higher reputation (a mechanical relationship). Fortunately, the opinion fixed-effect $\tau_p$ controls for all opinion characteristics, including the unobserved $U_{p}$, thus addressing potential confounding.

In summary, our baseline specifications address potential confounding due to (i) time-invariant challenger characteristics that affect persuasion, and (ii) users endogenously selecting which opinions to challenge.
In the next section, we introduce specifications that instrument for the challenger's reputation in each debate.
The instrumental variable specifications inherit the robustness of the baseline specifications to confounding from time-invariant challenger characteristics and endogenous opinion selection, while further addressing potential confounding due to time-varying challenger characteristics that affect debate success.

\subsection{Instrumental Variable Specifications}
\label{sec:instrument}

Our instrumental variable specifications address confounding due to unobserved user characteristics that affect persuasion and vary with their experience on the platform.
Estimates from this specification quantify the local average treatment effect (LATE) of reputation on debate success if instrument relevance, exogeneity, exclusion and monotonicity hold \parencite{imbens1994identification}. In this section, we derive our instrument and provide empirical evidence to support its validity.

Our instrument is motivated by the fact that a user's reputation largely depends on the number of posters persuaded previously, since other users who are not the poster rarely award $\Delta$s:
\begin{align*}
  r_{pu} \approx \sum_{p'<p} Y_{p'u}
\end{align*}
\indent From equation (\ref{eq:quality}), we also know that a user's position $t_{pu}$ in the sequence of challengers of opinion $p$ is correlated with debate success $Y_{pu}$. Hence, we define our instrument $Z_{pu}$ for the challenger's reputation $r_{pu}$ as the \textit{mean past position} of user $u$ before challenging opinion $p$:
\begin{eqnarray}
  Z_{pu} = \frac{\sum_{p'<p} t_{p'u} S_{p'u}}{\sum_{p'<p} S_{p'u}}
\end{eqnarray}
  where $p'$ is a chronologically-ordered opinion index, and the opinion selection indicator $S_{p'u}=1$ if user $u$ challenged opinion $p'$ and $S_{p'u}=0$ otherwise. We expect users who were consistently late challengers of opinions in the past (and thus, have larger mean past positions) to have persuaded fewer posters on average than users who were consistently early, and hence have lower reputation in the present. Thus, we expect $Z_{pu}$ to be negatively correlated with $r_{pu}$.

\noindent We confirm this relationship with the following first-stage regression:
\begin{align}
  r_{pu} &= \tau_p + a_1 Z_{pu} + a_2 s_{pu} + a_3 t_{pu} + \epsilon_{pu}
  \label{eq:livfs}
\end{align}
where $\tau_p$ is an opinion fixed-effect, $s_{pu}$ is user $u$'s skill, $t_{pu}$ is user $u$'s position in the sequence of challengers of opinion $p$ and $\epsilon_{pu}$ is a zero-mean Gaussian error term.

Our first-stage estimates in Table \ref{tab:firststage} indicate that a one unit increase in the mean past position of a user predicts a 0.18 unit decrease in their present reputation. The F-statistic on the instrument greatly exceeds the rule-of-thumb threshold \parencite{stock2005testing}, alleviating concerns about instrument strength. Skill has a positive first-stage correlation with reputation, which is expected since higher skilled users are likely to have persuaded more posters previously. The response position has a negative first-stage correlation with reputation, which we expect if users are consistent in their preference to respond early or late.

\addtolength{\tabcolsep}{14.5mm}
\begin{table*}[!t]
\vspace{5mm}
\small
\centering
\caption{\textbf{First-stage estimates.} Mean past position as an instrument for reputation.}
\begin{tabular}[t]{lc}
\toprule
& \multicolumn{1}{c}{Dependent Variable: Reputation $r_{pu}$}\\
\midrule
Mean past position $Z_{pu}$            & $-0.1833$ ($0.003$)$^{***}$                  \\
Skill $s_{pu}$ (percentage)            & $\phantom{-}2.3055$ ($0.012$)$^{***}$        \\
Position $t_{pu}$ (std. deviations)    & $-1.7354$ ($0.067$)$^{***}$                  \\
Opinion fixed-effects ($\tau_p$)       & \cmark                                       \\
Instrument F-Statistic                 & $3,338.7$                                      \\
No. of debates                         & $1,019,469$                                  \\
$R^2$                                  & $0.22$                                       \\
\bottomrule
\end{tabular}
\label{tab:firststage}
\vspace{3mm}

Note: Standard errors displayed in parentheses. $^{***}p<0.001; ^{**}p<0.01; ^{*}p<0.05$
\vspace{0mm}
\end{table*}%
\addtolength{\tabcolsep}{-14.5mm}

An immediate concern is users selecting opinions to challenge based on their \textit{anticipated position} in the sequence of challengers, since users can observe the number of ongoing and concluded debates with the poster before deciding to challenge an opinion. We characterize this scenario using the causal graph in Figure \ref{fig:sampleselection2}, which extends the causal graph in Figure \ref{fig:sampleselection} with shaded nodes $Z_{pu}$ (for the instrument) and $t_{pu}$ (for the challenger's present position). $t_{pu}$ affects the debate outcome $Y_{pu}$, based on equation (\ref{eq:quality}).
Recall from Section \ref{sec:baseline} that our specifications implicitly condition on the collider $S_{pu}=1$. If the instrument is correlated with opinion selection (depicted by the undirected edge $Z_{pu} \leftrightarrow S_{pu}$) and users select opinions to challenge based on their anticipated position (depicted by the edge $t_{pu} \rightarrow S_{pu}$), conditioning on $S_{pu}$ will open the confounding causal pathway $t_{pu} \rightarrow S_{pu} \leftrightarrow Z_{pu}$. Hence, it is essential to control for the challenger's present position $t_{pu}$, which could otherwise confound the instrument.

The causal graph in Figure \ref{fig:sampleselection2} reveals a second source of instrument confounding that has received recent attention \parencite{hughes2019selection,swanson2019practical}. If the instrument is correlated with opinion selection (depicted by the undirected edge $Z_{pu} \leftrightarrow S_{pu}$) and some unobserved opinion characteristic $U_{p}$  (such as the opinion topic) affects both opinion selection and debate success, conditioning on $S_{pu}$ opens the confounding causal pathway $U_{p} \rightarrow S_{pu} \leftrightarrow Z_{pu}$ that violates instrument exogeneity.

\noindent We can test for correlation between the instrument and opinion selection by estimating the following linear probability model of a user challenging more than one opinion after opinion $p$, where $\rho_u$ is a user fixed-effect, $m_{pu}$ is a calendar month-year fixed-effect and $\epsilon_{pu}$ is a Gaussian error term:
\begin{align*}
  \mathbb{I}[u \textrm{ challenges > 1 future opinion}] = \rho_u + m_{pu} + \theta_1 Z_{pu} + \theta_2 r_{pu} + \epsilon_{pu}
\end{align*}
The estimates of $\theta_1$ in Table \ref{tab:participation2} suggest a small but significant negative correlation between $Z_{pu}$ and $S_{pu}$, justifying our concerns of endogenous opinion selection violating instrument exogeneity. Fortunately (as discussed Section \ref{sec:baseline}), the opinion fixed-effect $\tau_p$ controls for all opinion characteristics, including unobserved $U_{p}$. This alleviates concerns of instrument exogeneity being violated due to endogenous opinion selection.

\begin{minipage}{\linewidth}
  \vspace{-5mm}
  \hspace{-9mm}
  \begin{minipage}[b]{0.522\linewidth}
    \centering
    \includegraphics[width=2.75in]{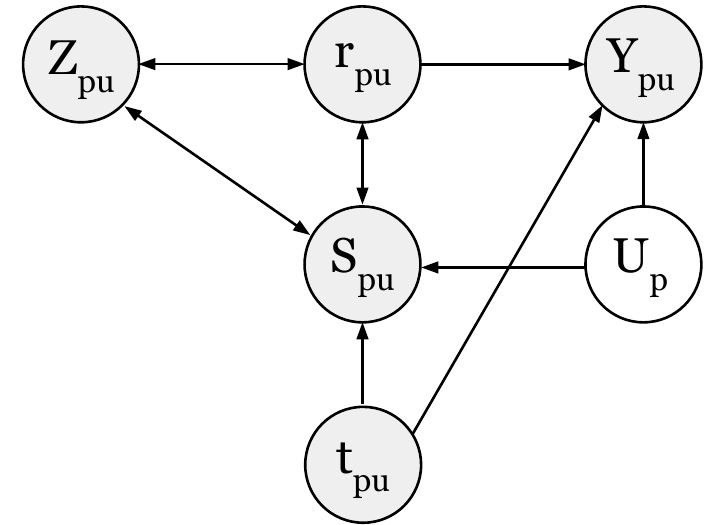}
    \captionof{figure}{\textbf{Instrument confounding via opinion selection.}}
    \label{fig:sampleselection2}
  \end{minipage}
  \begin{minipage}[b]{0.55\linewidth}
    \small
    \centering
    \begin{tabular}[t]{lc}
    \toprule
    \multicolumn{2}{c}{
      Dependent Variable: $\mathbb{I}[u \textrm{ challenges > 1 future opinion}]$
      }\\
    \midrule
    Mean past position $Z_{pu}$                   & $-0.0040$ ($0.00003$)$^{***}$ \\
    Reputation $r_{pu}$ (10 units)                & $\phantom{-}0.0166$ ($0.00012$)$^{***}$ \\
    User fixed-effects ($\rho_u$)                 & \cmark \\
    Month-year fixed-effects ($m_{pu}$)           & \cmark \\
    No. of debates                                & $947,181$ \\
    $R^2$                                         & $0.57$    \\
    \bottomrule
    \end{tabular}
    \label{tab:sample-selection-instrument}
    \vspace{3mm}

    Note: Standard errors displayed in parentheses.

    $^{***}p<0.001; ^{**}p<0.01; ^{*}p<0.05$

    \captionof{table}{\textbf{Instrument/opinion selection correlation.}}
    \label{tab:participation2}
    \end{minipage}
    \vspace{-2.5mm}
\end{minipage}

Another plausible concern is of the instrument affecting the debate outcome via channels that do not include the user's reputation, which violates the instrument exclusion restriction. For example, if users learn to be more persuasive from the earlier challengers of an opinion, a user with a high mean past position could be more persuasive in the present than one with a low mean past position.

\begin{wrapfigure}[11]{r}{0.28\textwidth}
  \vspace{-5mm}
  \includegraphics[width=0.975\linewidth,right]{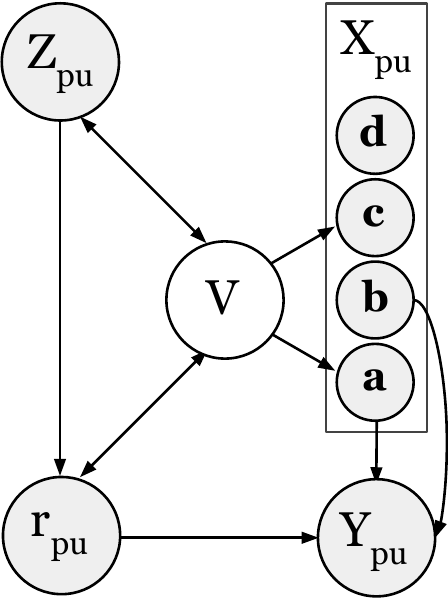}
  \label{fig:dag-intuition}
\end{wrapfigure}
We address this concern in two ways. First, note that any user characteristic correlated with successful persuasion is likely to affect the debate outcome through the text of their responses. Hence, controlling for the response text will block direct channels of influence between the instrument and the debate outcome. This is formalized by the causal graph on the right. Here, the reputation $r_{pu}$, debate outcome $Y_{pu}$, response text $X_{pu}$ and instrument $Z_{pu}$ are observed. $V$ contains all unobserved confounders of the instrument or reputation (or both) that affect the outcome through the text $X_{pu}$.
If we decompose the text into conceptual components \textbf{a}, \textbf{b}, \textbf{c} and \textbf{d}, it is sufficient to control for \textbf{a} to block the $Z_{pu} \leftrightarrow V \rightarrow \pmb{a} \rightarrow Y_{pu}$ causal pathway.

\noindent We operationalize this idea by estimating the following partially-linear instrumental variable specification
with endogenous $r_{pu}$, as formulated by \parencite{chernozhukov2018double}:
\begin{align*}
  Y_{pu} &=  \beta_1 r_{pu} + \beta_2 s_{pu} + \beta_3 t_{pu} + g(\tau_p, X_{pu}) + \epsilon_{pu}
         &\mathbb{E}[\epsilon_{pu}| Z_{pu}, \tau_p, s_{pu}, t_{pu}, X_{pu}] = 0 \nonumber\\
  Z_{pu} &=  \alpha_1 s_{pu} + \alpha_2 t_{pu} + h(\tau_p, X_{pu}) + \epsilon'_{pu}
         &\mathbb{E}[\epsilon'_{pu}|\tau_p, s_{pu}, t_{pu}, X_{pu}] = 0
  \label{eq:pliv}
\end{align*}
In this specification, the high-dimensional covariates $\tau_p$ (the opinion fixed-effects) and $X_{pu}$ (a vector representation of $u$'s response text) have been moved into the arguments of the ``nuisance functions'' $g(\cdot)$ and $h(\cdot)$.
As earlier, $r_{pu}$ is $u$'s reputation, $s_{pu}$ is $u$'s skill, $t_{pu}$ is $u$'s position and $Z_{pu}$ (the instrument) is the mean past position of $u$ before opinion $p$. $\epsilon_{pu}$ and $\epsilon'_{pu}$ are error terms with zero conditional mean.  $\beta_1$ is the parameter of interest, quantifying the causal effect of reputation on persuasion.

No distributional assumptions are placed on $\epsilon_{pu}$ and $\epsilon'_{pu}$, and hence, this specification does not assume any functional form (in contrast with logit, probit and linear probability models). $g(\cdot)$ and $h(\cdot)$ can be flexible nonparametric functions. We discuss estimation and inference in Section \ref{sec:estimation}.

Second, we use the ``plausibly exogenous'' instrumental variable framework \parencite{conley2012plausibly} to relax the instrument exclusion restriction and include $Z_{pu}$ directly in the debate outcome model\footnote{\parencite{conley2012plausibly} proposes four inference strategies that incorporate plausibly exogenous instruments. The inference strategy we use relies on the fewest assumptions and provides the most conservative estimates of $\beta_1$.}:
\begin{align*}
  Y_{pu} &= \tau_p  + \beta_1 r_{pu} + \beta_2 s_{pu} + \beta_3 t_{pu} + \gamma Z_{pu}  + \epsilon_{pu}
         &\mathbb{E}[\epsilon_{pu}| Z_{pu}, \tau_p, s_{pu}, t_{pu}] = 0 \nonumber\\
  Z_{pu} &= \tau_p  + \alpha_1 s_{pu} + \alpha_2 t_{pu} + \epsilon'_{pu}
         &\mathbb{E}[\epsilon'_{pu}|\tau_p, s_{pu}, t_{pu}] = 0
\end{align*}
where $\gamma \neq 0$ encodes by how much the exclusion restriction is violated. For a fixed $\gamma$ and conditionally exogenous instrument, the effect of reputation on debate success can be quantified via two-stage least-squares estimation of the following regression, using $Z_{pu}$ as an instrument for $r_{pu}$:
\begin{align*}
  (Y_{pu} - \gamma Z_{pu})
    &= \tau_p  + \beta_1 r_{pu} + \beta_2 s_{pu} + \beta_3 t_{pu}  + \epsilon_{pu} \nonumber
\end{align*}
If users indeed learn to be more persuasive from earlier challengers, we would expect $\gamma > 0$. We report estimates of $\beta_1$ from the specification above for a range of $\gamma$ values in Section \ref{sec:results}.

While instrument relevance, exogeneity and exclusion are sufficient to guarantee identification of the effect of reputation on debate success, we also require instrument monotonicity to interpret our estimate as a local average treatment effect (LATE) \parencite{imbens1994identification}.
Instrument monotonicity will be violated if there exists a subpopulation of debates where increasing the mean past position of the challenger would \textit{increase} their present reputation, and decreasing their mean past position would \textit{decrease} their present reputation (members of this subpopulation are called \textit{defiers}). Such challengers are \textit{more} likely to persuade a poster when they respond later. The large and precisely-estimated negative within-user correlation between the number of earlier challengers and debate success ($\hat{\theta_2} = -0.0107 \pm 0.0003$) in Table \ref{tab:experience} suggests that the existence of such challengers is unlikely.

The LATE is the effect of reputation on debate success for \textit{compliers}, comprised of debates where the challenger's reputation is indeed affected by their mean past position. The challengers in these debate subpopulations are more persuasive at earlier (lower) positions, and less persuasive at later (higher) positions. Hence, we expect that the compliers exclude debates with challengers having low persuasive ability, who are unlikely to be more or less persuasive in any position. We also expect challengers with moderate to high persuasive ability to benefit more from an increase in their reputation than challengers with low persuasive ability, since a high reputation is unlikely to substitute for low persuasive ability. Hence, we expect the LATE to be larger than the average treatment effect of reputation on debate success.

\subsection{Estimation and Inference}
\label{sec:estimation}

Our baseline linear probability model and linear instrumental variable specifications can be estimated using ordinary least-squares and two-stage least squares respectively, adapted to accommodate high-dimensional fixed-effects \parencite{correia2017hdfe}. In this section, we describe how the double machine-learning framework \parencite{chernozhukov2018double} can be used to consistently estimate the effects of reputation, skill and position in the partially-linear instrumental variable specification.

Double machine-learning extends the partialling-out procedure of Frisch-Waugh-Lovell \parencite{frisch1933partial,lovell1963seasonal} to use flexible nonparametric functions estimated via machine learning. We first describe the basic setup assuming reputation is conditionally exogenous (given the response text), ignoring the opinion fixed-effects, and ignoring the effects of skill and position. Consider the following partially-linear probability model:
\begin{align*}
  Y_{pu} &= \beta r_{pu} + g(X_{pu}) + \epsilon_{pu} &\mathbb{E}[\epsilon_{pu}| r_{pu}, X_{pu}] = 0 \nonumber\\
  r_{pu} &= h(X_{pu}) + \epsilon'_{pu}               &\mathbb{E}[\epsilon'_{pu}| X_{pu}] = 0
\end{align*}
where $Y_{pu}$ is the debate outcome, $r_{pu}$ is the challenger's reputation, $X_{pu}$ is their response text and $\epsilon_{pu}$ and $\epsilon'_{pu}$ are Gaussian error terms with zero conditional mean.  $g(\cdot)$ and $h(\cdot)$ are unknown nonparametric functions. We are interested in consistently estimating and performing valid inference on $\beta$.

If $g(\cdot)$ and $h(\cdot)$ were fixed and known, consistent estimation is possible by solving for $\beta$ in an empirical version of the following moment condition (equivalent to ordinary least-squares estimation):
\begin{align*}
  \mathbb{E}[(Y_{pu} - g(X_{pu}) - \beta r_{pu})r_{pu}] = 0
\end{align*}
However, $g(\cdot)$ is unknown and needs to be jointly estimated with $\beta$. A solution is to first estimate $g(\cdot)$ on a separate subsample $\mathcal{S}'$ of the data, and then estimate $\beta$ by solving an empirical version of the moment condition above on the remaining subsample $\mathcal{S}$. This procedure, called \textit{sample-splitting}, eliminates the ``overfitting-bias'' introduced in the process of estimating $g(\cdot)$.

If $g(\cdot)$ is estimated via machine learning, the procedure above results in inconsistent estimates $\hat{\beta}$. \parencite{chernozhukov2018double} decomposes the scaled bias of $\hat{\beta}$ into the following two terms:
\begin{align*}
  \sqrt{n}(\hat{\beta} - \beta)
    = \underbrace{
        ( \frac{1}{n} \sum_{(p,u) \in \mathcal{S}} r_{pu}^2 )^{-1} \frac{1}{\sqrt{n}} \sum_{(p,u) \in \mathcal{S}} r_{pu} \epsilon_{pu}
      }_{\textrm{Term } a} +
      \underbrace{
        ( \frac{1}{n} \sum_{(p,u) \in \mathcal{S}} r_{pu}^2 )^{-1}
        \frac{1}{\sqrt{n}} \sum_{(p,u) \in \mathcal{S}} r_{pu} (g(X_{pu}) - \hat{g}(X_{pu}))
      }_{\textrm{Term } b}
\end{align*}
Term $a$ converges at a $n^{-1/2}$ rate to a zero-mean Gaussian. However, by virtue of $g(\cdot)$ being estimated via machine learning, term $b$ will typically converge to zero at a rate slower than $n^{-1/2}$ due to the slow convergence of the estimation error $g(X_{pu}) - \hat{g}(X_{pu})$. This is called the ``regularization bias'' of $\hat{g}(\cdot)$.

Double machine-learning eliminates regularization bias via a procedure called orthogonalization. $\beta$ is estimated by solving an empirical version of the following ``Neyman-orthogonal'' moment condition:
\begin{align*}
  \mathbb{E}[( (Y_{pu} - \mathbb{E}[Y_{pu}|X_{pu}]) -
               \beta(r_{pu} - \mathbb{E}[r_{pu}|X_{pu}]))(r_{pu} - \mathbb{E}[r_{pu}|X_{pu}])] = 0
\end{align*}
The empirical version of this moment condition can be solved via a procedure similar to the residual-on-residuals regression of \parencite{robinson1988root}. The procedure is as follows (where $\mathcal{S}$ and $\mathcal{S}'$ are disjoint subsamples of the data, and $m(\cdot)$ and $l(\cdot)$ are nonparametric functions):
\begin{enumerate}[itemsep=2mm]
  \item Estimate the conditional expectation function $l(X_{pu}) = \mathbb{E}[Y_{pu}|X_{pu}]$ on $\mathcal{S}'$ to get $\hat{l}(\cdot)$.
  \item Estimate the conditional expectation function $m(X_{pu}) = \mathbb{E}[r_{pu}|X_{pu}]$ on $\mathcal{S}'$ to get $\hat{m}(\cdot)$.
  \item Estimate the outcome residual $\tilde{Y}_{pu} = Y_{pu} - \hat{l}(X_{pu})$ on $\mathcal{S}$.
  \item Estimate the treatment residual $\tilde{r}_{pu} = r_{pu} - \hat{m}(X_{pu})$ on $\mathcal{S}$.
  \item Regress $\tilde{Y}_{pu}$ on $\tilde{r}_{pu}$ to obtain $\hat{\beta}$.
\end{enumerate}
Note that we no longer need to estimate $g(\cdot)$, and instead need to estimate the conditional expectations $l(\cdot)$ and $m(\cdot)$ that can be arbitrary nonparametric functions of $X_{pu}$ (such as neural networks). This procedure can be extended to include skill and position as controls by estimating additional conditional expectation functions to predict the challenger's skill and position from their response text on $S'$, estimating the residuals $\tilde{s}_{pu}$ and $\tilde{t}_{pu}$ on $S$, and then regressing $\tilde{Y}_{pu}$ on $\tilde{r}_{pu}$, $\tilde{s}_{pu}$ and $\tilde{t}_{pu}$.

The resulting estimate $\hat{\beta}$ is $\sqrt{n}$-consistent and asymptotically normal. \parencite{chernozhukov2018double} shows that term $b$ of the scaled bias of $\hat{\beta}$ is now given by following expression:
\begin{align*}
  (\frac{1}{n} \sum_{(p,u) \in \mathcal{S}} V_{pu}^2 )^{-1}
  \frac{1}{\sqrt{n}}
    \sum_{(p,u) \in \mathcal{S}}
      \underbrace{(m(X_{pu}) - \hat{m}(X_{pu}))}_{\hat{m}(\cdot) \textrm{estimation error}}
      \underbrace{(l(X_{pu}) - \hat{l}(X_{pu}))}_{\hat{l}(\cdot)  \textrm{estimation error}}
\end{align*}
This contains the product of nuisance function estimation errors. Hence, orthogonalization enables $\sqrt{n}$-consistent estimation of $\hat{\beta}$ as long as the \textit{product} of the convergence rates of $\hat{m}(\cdot)$ and $\hat{l}(\cdot)$ is $n^{-1/2}$. This is more viable than requiring \textit{each} nuisance function to converge at a $n^{-1/2}$ rate.

If $r_{pu}$ is endogenous and $Z_{pu}$ is a valid instrument for $r_{pu}$, (Chernozhukov et al., 2018) proposes the following Neyman-orthogonal moment condition to estimate $\beta$ in a partially-linear instrumental variable specification:
\begin{align}
  \mathbb{E}[( (Y_{pu} - \mathbb{E}[Y_{pu}|X_{pu}]) -
               \beta(r_{pu} - \mathbb{E}[r_{pu}|X_{pu}]))(Z_{pu} - \mathbb{E}[Z_{pu}|X_{pu}])] = 0
\end{align}
By a similar bias derivation, the estimated $\hat{\beta}$ is shown to be $\sqrt{n}$-consistent and asymptotically normal, as long
as the instrument is valid and the product of the nuisance function convergence rates is $n^{-1/2}$.

We now detail our overall estimation procedure for the partially-linear instrumental variable specification. We include the opinion fixed-effect $\tau_p$, skill $s_{pu}$ and position $t_{pu}$ as controls. $\mathcal{S}$ and $\mathcal{S}'$ are disjoint subsamples of the data, and $m_r(\cdot), m_s(\cdot), m_t(\cdot), m_p(\cdot)$, $l(\cdot)$ and $q(\cdot)$ are nonparametric functions that we detail in the next subsection. The procedure is as follows:
\begin{enumerate}[itemsep=0pt]
  \item Estimate the following conditional expectation functions on sample $\mathcal{S}'$:
  \begin{multicols}{2}
    \begin{enumerate}[label=\roman*.]
      \item $l(X_{pu}, \tau_p) = \mathbb{E}[Y_{pu}|X_{pu}, \tau_p]$  to get $\hat{l}(\cdot)$.
      \item $q(X_{pu}, \tau_p) = \mathbb{E}[Z_{pu}|X_{pu}, \tau_p]$ to get $\hat{q}(\cdot)$.
      \item[]
      \item $m_r(X_{pu}, \tau_p) = \mathbb{E}[r_{pu}|X_{pu}, \tau_p]$ to get $\hat{m}_r(\cdot)$.
      \item $m_s(X_{pu}, \tau_p) = \mathbb{E}[s_{pu}|X_{pu}, \tau_p]$ to get $\hat{m}_s(\cdot)$.
      \item $m_t(X_{pu}, \tau_p) = \mathbb{E}[t_{pu}|X_{pu}, \tau_p]$ to get $\hat{m}_t(\cdot)$.
    \end{enumerate}
  \end{multicols}
  \item Estimate the following residuals on sample $\mathcal{S}$:
  \begin{multicols}{2}
    \begin{enumerate}[label=\roman*.]
      \item $\tilde{Y}_{pu} = Y_{pu} - \hat{l}(X_{pu}, \tau_p)$.
      \item $\tilde{Z}_{pu} = Z_{pu} - \hat{q}(X_{pu}, \tau_p)$.
      \item[]
      \item $\tilde{r}_{pu} = r_{pu} - \hat{m}_r(X_{pu}, \tau_p)$.
      \item $\tilde{s}_{pu} = s_{pu} - \hat{m}_s(X_{pu}, \tau_p)$.
      \item $\tilde{t}_{pu} = t_{pu} - \hat{m}_t(X_{pu}, \tau_p)$.
    \end{enumerate}
  \end{multicols}
  \item Run a two-stage least-squares regression of $\tilde{Y}_{pu}$ on $\tilde{r}_{pu}, \tilde{s}_{pu}, \tilde{t}_{pu}$ using $\tilde{Z}_{pu}$ as an instrument for $\tilde{r}_{pu}$ to obtain the estimated local average treatment effects of reputation, skill and position on debate success.
\end{enumerate}

We partition the debates for opinions with more than one response (mirroring the data used in the specifications with opinion fixed-effects) uniformly at random into an estimation subsample $S'$ containing 10\% of the debates (101,946 debates) and an inference subsample $S$ containing 90\% of the debates (917,523 debates), ensuring that every opinion is represented in both $S$ and $S'$. In the next section, we describe how we use neural networks with rectified linear unit (ReLU) activation functions for the nonparametric functions $m_r(\cdot), m_s(\cdot), m_t(\cdot), m_p(\cdot)$, $l(\cdot)$ and $q(\cdot)$, which have been shown to converge at $n^{-1/4}$ rates \parencite{farrell2018deep} that enables consistent estimation and valid inference.

\subsection{Neural Models of Text as Nuisance Functions}
\label{sec:ml}

A fully-connected neural network with $h$ hidden layers is parameterized by matrices $\pmb{W}_1, \dots, \pmb{W}_{h+1}$ and \textit{activation functions} (called activations) $a_1, \dots, a_{h+1}$. The hidden layer \textit{sizes} $s_1, \dots, s_{h}$ are architectural hyperparameters that determine the sizes of the matrices $\pmb{W}_1, \dots, \pmb{W}_{h+1}$ as follows, where $D$ and $O$ are the dimensionalities of the neural network input and output, respectively:
\begin{align*}
  &\textrm{Size of } \pmb{W}_1 = D \times s_1\\
  &\textrm{Size of } \pmb{W}_i = s_{i-1} \times s_{i} \textrm{ for } i = 2, \dots, h\\
  &\textrm{Size of } \pmb{W}_{h+1} = s_{h} \times O
\end{align*}

\begin{figure*}[!t]
  \vspace{-4mm}
  \centering
  \includegraphics[width=0.99\linewidth]{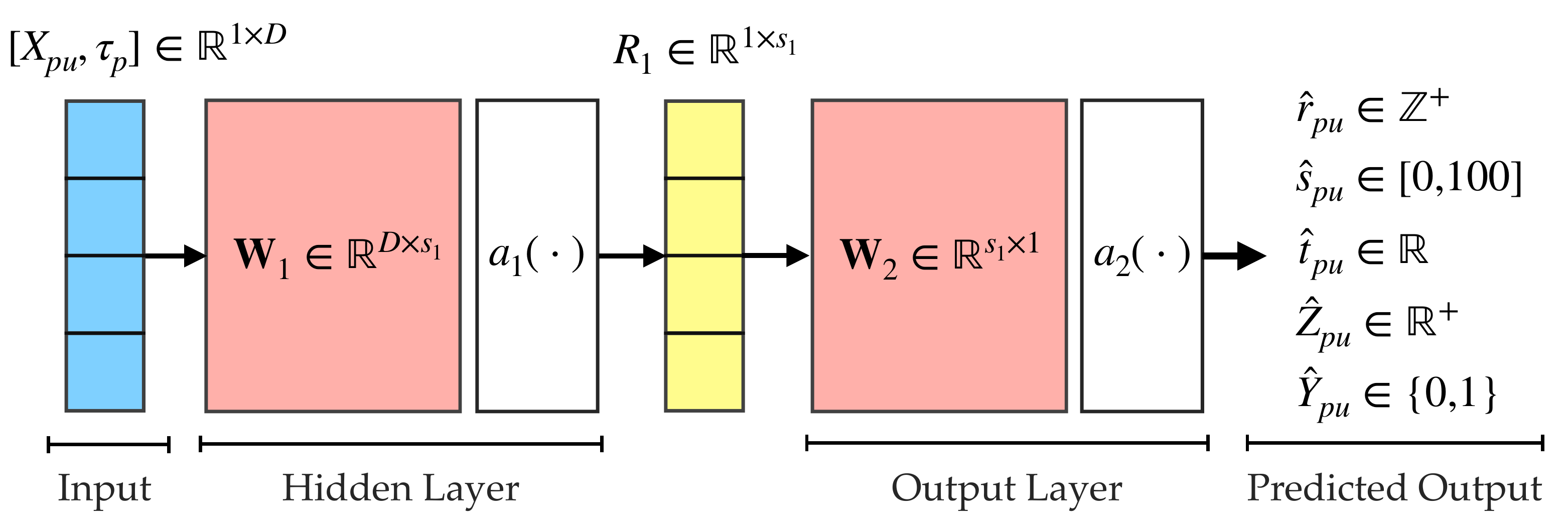}
  \caption{\textbf{A neural network with one hidden layer ($h=1$).} The neural network transforms the $D$-dimensional input, a concatenation of the response text vector $X_{pu}$ and the fixed-effects indicator vector for $\tau_p$, into a 1-dimensional output. $\pmb{W}_1$ and $\pmb{W}_2$ are parameters to be estimated. $a_1(\cdot)$ and $a_2(\cdot)$ are activation functions.}
  \label{fig:dnn}
\end{figure*}

Each layer $i$ multiplies the intermediate vector $R_{i-1}$ produced by the previous layer with $\pmb{W}_i$, and applies the activation function $a_i(\cdot)$ to produce $R_{i} = a_i(R_{i-1}\pmb{W}_i)$. Figure \ref{fig:dnn} illustrates a neural network with one hidden layer ($h=1$), input dimensionality $D$ and output dimensionality $O=1$. The neural network transforms the input, a concatenation of the response text vector $X_{pu}$ and the fixed-effects indicator vector for $\tau_p$, into the 1-dimensional predicted output $a_2(a_1([X_{pu}, \tau_p] \times \pmb{W}_1 ) \times \pmb{W}_2)$.

We estimate five neural networks with rectified linear unit (ReLU) activations to predict (i) debate success $Y_{pu} \in \{0,1\}$, (ii) reputation $r_{pu} \in \mathbb{Z}^+$, (iii) skill $s_{pu} \in [0, 100]$ (as a percentage), (iv) position $t_{pu} \in \mathbb{R}$ (standardized to have zero-mean and unit-variance) and (v) the instrument $Z_{pu} \in \mathbb{R}^+$ from the response text $X_{pu}$ and opinion fixed-effects $\tau_p$. Though recurrent \parencite{hochreiter1997long} and convolutional \parencite{kim2014convolutional} neural networks are more popular for textual prediction tasks, ReLU neural networks have guaranteed $n^{-1/4}$ convergence rates \parencite{farrell2018deep} that we require for consistent estimation and valid inference. Hence, we set each of the hidden layer activations $a_1(\cdot), \dots, a_{h}(\cdot)$ to the rectifier function $a_i(x) = \textrm{max}(0, x)$. Since the output of each neural network is one-dimensional, we set the size of the output layer matrix $\pmb{W}_{h+1}$ to $s_{h} \times 1$.

\noindent \textbf{Output layer activations and loss functions.} For the debate success prediction network with the binary target $Y_{pu} \in \{0,1\}$, we set the output layer activation to the logistic sigmoid function: $a_{h+1}(x) = (1+e^{-x})^{-1} \in [0,1]$. For the skill prediction network with the bounded target $s_{pu} \in [0,100]$, we set the output layer activation to the scaled logistic sigmoid function: $a_{h+1}(x) = (1+e^{-x})^{-1} \times 100 \in [0,100]$. For the reputation and instrument prediction networks with nonnegative targets $r_{pu} \in \mathbb{Z}^+$ and $Z_{pu} \in \mathbb{R}^+$, we set the output layer activation to the rectifier function: $a_{h+1}(x) = \textrm{max}(0, x)$. For the position prediction network with unbounded target $t_{pu} \in \mathbb{R}$, we set the output layer activation to the identity function: $a_{h+1}(x) = x$. We estimate the parameters $\pmb{W}_1, \dots, \pmb{W}_{h+1}$ for the debate success prediction network by minimizing the binary cross-entropy loss $Y_{pu}\textrm{log}(\hat{Y}_{pu}) + (1-Y_{pu})\textrm{log}(1-\hat{Y}_{pu})$ (where $\hat{Y}_{pu}$ is the predicted output), and for the other networks by minimizing the mean squared error.

\noindent \textbf{Neural network input.} We follow recommendations from the text-as-data literature \parencite{gentzkow2019text} and construct a term-frequency inverse-document-frequency (TF-IDF) matrix from the text of the challengers' responses. We preprocess the text to remove links, numbers, pronouns, punctuation and text formatting symbols, and replace each word with its lower-cased stem (for example, ``economically'' and ``economics'' will be replaced by the stem ``economic''). We exclude very rare words (present in less than 0.1\% of the responses) and very frequent words (present in more than 99.9\% of the responses), since these words will contribute negligibly towards more accurate predictions. The final vocabulary contains 4,926 distinct words. Each row of the TF-IDF matrix corresponds to a vector $X_{pu} \in \mathbb{R}^{4926}$. We also construct an indicator vector $\tau_p \in \{0,1\}^{84998}$ (since there are 84,998 unique opinion clusters), where only the $p^{\textrm{th}}$ element of $\tau_p$ is set to 1 and the rest are set to zero. The concatenation of these two vectors, $[X_{pu}, \tau_p] \in \mathbb{R}^{89924}$, is passed as input to the neural networks.

\noindent \textbf{Optimization.} We train each network via backpropagation \parencite{rumelhart1986learning} with the Adam gradient-based optimization algorithm \parencite{kingma2015adam} iterating over mini-batches of the training data. We begin the optimization process by initializing the parameters using the \textit{Kaiming uniform} initialization scheme \parencite{he2015delving}, which has been shown to perform well both empirically and theoretically \parencite{hanin2018start}. We perform batch-normalization \parencite{ioffe2015batch} on each layer's output after applying the activation function to prevent internal covariate shift and accelerate convergence. To prevent overfitting to the training data, we apply weight-decay (a form of $L_2$-norm penalization) \parencite{krogh1992simple} to all the parameters, along with early-stopping (halting the training process once the out-of-sample predictive power starts decreasing with training iterations). We do not employ dropout regularization \parencite{srivastava2014dropout}, since it reduces out-of-sample predictive power when combined with batch-normalization \parencite{li2019understanding}.

\noindent \textbf{Architectural and optimization hyperparameters.} The number of hidden layers $h$, hidden layer sizes $s_1, \dots, s_{h}$, weight-decay penalty, optimization learning rate and mini-batch size are architectural and optimization hyperparameters that need to be tuned empirically.
Hence, we further partition the debates in the estimation subsample $S'$ uniformly at random into a training subsample $S'_{\textrm{train}}$ containing 75\% of the debates (76,459 debates) and a validation subsample $S'_{\textrm{val}}$ containing 25\% of the debates (25,487 debates). During the
hyperparameter tuning process, we train the neural network on $S'_{\textrm{train}}$ and evaluate its loss at each training iteration on both $S'_{\textrm{train}}$ and $S'_{\textrm{val}}$.

We fix the size of the hidden layers $s_1, \dots, s_{h}$ to the dimensionality of the response text vector $X_{pu}$ (=4,926) and tune the number of hidden layers for each neural network. Deep, fixed-width ReLU networks of this type have been shown to generalize well both empirically and theoretically \parencite{safran2017depth,hanin2019universal}. For each neural network, we evaluate the \text{training} loss (for at most 5,000 mini-batch iterations with early-stopping) with an increasing number of hidden layers, until the training loss no longer improves. Each neural network with the number of hidden layers thus found has enough representational capacity to capture patterns in the training data, but is likely to have overfit the training data and suffer from poor out-of-sample predictive power.

\addtolength{\tabcolsep}{0.2mm}
\begin{table*}[!h]
\small
\centering
\begin{tabular}[t]{lcccl}
\toprule
& Number of
& \multicolumn{2}{c}{Activation Functions}
& %
\\
\cmidrule(lr){3-4}
Prediction target
& Hidden layers  %
& \text{Hidden Layer} %
& \text{Output Layer} %
& Loss Function
\\
\midrule
Debate success $Y_{pu} \in \{0,1\}$             & 5  & ReLU & Sigmoid   & Binary Cross-Entropy\\[1.5mm]
Reputation $r_{pu} \in \mathbb{Z}^+$            & 3  & ReLU & Rectifier & Mean squared error\\[1.5mm]
Skill $s_{pu} \in [0,100]$ (percentage)         & 3  & ReLU & Sigmoid   & Mean squared error\\[1.5mm]
Position $t_{pu} \in \mathbb{R}$ (standardized) & 3  & ReLU & Identity  & Mean squared error\\[1.5mm]
Instrument $Z_{pu} \in \mathbb{R}^+$            & 5  & ReLU & Rectifier & Mean squared error\\[0mm]
\bottomrule
\end{tabular}
\caption{\textbf{Architectural hyperparameters.} The input layer matrix $\pmb{W}_1$ of each neural network has size 89,924 $\times$ 4,926, where 89,924 is the dimensionality of the input vector (the vocabulary size + the number of unique opinion clusters) and 4,926 is the dimensionality of $X_{pu}$ (the vocabulary size). Each of the $h$ hidden layer matrices $\pmb{W}_{2}, \dots \pmb{W}_{h}$ has size 4,926 $\times$ 4,926, and the output layer matrix $\pmb{W}_{h+1}$ has size 4,926 $\times$ 1.}
\label{tab:networks}
\end{table*}%
\addtolength{\tabcolsep}{-0.2mm}

\addtolength{\tabcolsep}{-0.6mm}
\begin{table*}[!h]
\small
\centering
\begin{tabular}[t]{lcccccc}
\toprule
                  & & & & \multicolumn{3}{c}{Subsample Loss}\\
                                                    \cmidrule(lr){5-7}
Prediction target & Learning Rate & Batch Size & Weight-Decay            & Train & Validation & Inference\\
\midrule
Debate success $Y_{pu} \in \{0,1\}$             & 0.0001 & 50,000 & 10000             & \phantom{1}0.148  & \phantom{1}0.155  & \phantom{1}0.152 \\[1.5mm]
Reputation $r_{pu} \in \mathbb{Z}^+$            & 0.0001 & 50,000 & \phantom{000}10   & 39.801            & 40.406             & 39.842          \\[1.5mm]
Skill $s_{pu} \in [0,100]$ (percentage)         & 0.0001 & 50,000 & \phantom{000}10   & \phantom{1}3.672  & \phantom{1}3.764  & \phantom{1}3.707 \\[1.5mm]
Position $t_{pu} \in \mathbb{R}$ (standardized) & 0.0001 & 50,000 & \phantom{000}10   & \phantom{1}0.658  & \phantom{1}0.789  & \phantom{1}0.796 \\[1.5mm]
Instrument $Z_{pu} \in \mathbb{R}^+$            & 0.0001 & 50,000 & 10000             & 12.389 & 13.370 & 13.217 \\[0mm]
\bottomrule
\end{tabular}
\caption{\textbf{Optimization hyperparameters.} The subsample losses on $S'_{\textrm{train}}$, $S'_{\textrm{val}}$ and $S$ are reported after training each neural network with the selected hyperparameters for at most 5,000 mini-batch iterations (with early-stopping) on $S'_{\textrm{train}}$.
The binary cross-entropy subsample loss is reported for the network predicting $Y_{pu}$ and the \textit{root} mean squared prediction error is reported for the other networks.}
\label{tab:networks2}
\end{table*}%
\addtolength{\tabcolsep}{0.6mm}
~\\[-7.5mm]
Hence, after having selected the number of hidden layers for each neural network via the aforementioned procedure, we evaluate the \textit{validation} loss of each neural network (for at most 5,000 mini-batch iterations with early-stopping) with an increasingly large weight-decay penalty (in the logarithmically-spaced range $0.001, 0.01, 0.1, \dots$), until the validation loss no longer improves. The final neural network thus found will have sufficient representational capacity and be sufficiently regularized to generalize well out-of-sample. During the process of tuning the number of hidden layers and the weight-decay penalty, we also empirically evaluate and select the values of the learning rate and mini-batch size that deliver the minimum validation loss with fast and stable convergence.

Table \ref{tab:networks} summarizes the selected architectural hyperparameters. Table \ref{tab:networks2} summarizes the selected optimization hyperparameters and the losses on each data subsample, which reflect the extent to which each target is correlated with potential confounders present in response text. After fixing the selected hyperparameters, we re-estimate the neural networks with on the full estimation subsample $S'$, estimate the prediction residuals on the inference sample $S$ and run a two-stage least-squares regression with these residuals, as described in the double machine-learning procedure in Section \ref{sec:estimation}.

\clearpage

\section{Results}
\label{sec:results}
\addtolength{\tabcolsep}{1mm}
\begin{table*}[!h]
\small
\centering
\caption{\textbf{Main results.} Estimated effects of reputation, skill and position on debate success with a logit model without opinion fixed-effects (1), linear probability models without (2) and with (3) opinion fixed-effects, linear instrumental variable (4) and partially-linear instrumental variable (5) specifications. Position is standardized to have zero-mean and unit-variance. Average marginal effects and pseudo-$R^2$ are reported for the logit model. The instrument $Z_{pu}$ is the mean past position of user $u$ before they challenged opinion $p$.}
\begin{tabular}[t]{lccccc}
\toprule
& \multicolumn{5}{c}{Dependent Variable: Debate Success $Y_{pu}$}\\
\midrule
& (1) & (2) & (3) & (4) & (5)\\
\midrule

Reputation $r_{pu}$ (10 units)
& $\phantom{-}0.0006$$^{***}$
& $\phantom{-}0.0010$$^{***}$
& $\phantom{-}0.0014$$^{***}$
& $\phantom{-}0.0109$$^{***}$
& $\phantom{-}0.0091$$^{***}$ \\
& ($0.00003$) & ($0.00006$)
& ($0.00006$) & ($0.00059$)
& ($0.00079$) \\[2mm]

Skill $s_{pu}$ (percentage)
& $\phantom{-}0.0024$$^{***}$
& $\phantom{-}0.0040$$^{***}$
& $\phantom{-}0.0035$$^{***}$
& $\phantom{-}0.0012$$^{***}$
& $\phantom{-}0.0016$$^{***}$ \\
& ($0.00004$) & ($0.00007$)
& ($0.00007$) & ($0.00016$)
& ($0.00024$) \\[2mm]

Position $t_{pu}$ (std. deviations)
& $-0.0479$$^{***}$ & $-0.0093$$^{***}$
& $-0.0097$$^{***}$ & $-0.0078$$^{***}$
& $-0.0088$$^{***}$ \\
& ($0.00265$) & ($0.00078$)
& ($0.00087$) & ($0.00075$)
& ($0.00077$) \\[2mm]

Response text ($X_{pu}$)               & \xmark      & \xmark       & \xmark      & \xmark      & \cmark      \\[1mm]
Instrument ($Z_{pu}$)                  & \xmark      & \xmark       & \xmark      & \cmark      & \cmark      \\[1mm]
Opinion fixed-effects ($\tau_p$)       & \xmark      & \xmark       & \cmark      & \cmark      & \cmark      \\[1mm]
No. of debates                         & $1,026,201$ & $1,026,201$  & $1,019,469$ & $1,019,469$ & $1,019,469$ \\[1mm]
$R^2$                                  & $0.051$     & $0.012$      & $0.203$     & ---         & --- \\
\bottomrule
\end{tabular}
\label{tab:mainresults}
\vspace{3mm}

Note: Standard errors (clustered by opinion) displayed in parentheses. $^{***}p<0.001; ^{**}p<0.01; ^{*}p<0.05$
\vspace{0mm}
\end{table*}%
\addtolength{\tabcolsep}{-1mm}

Table \ref{tab:mainresults} reports the estimated (marginal) effects of reputation, skill and position on debate success.
We exclude opinion fixed-effects from the logit model to prevent dropping debates for opinions where none of the challengers persuaded the poster, which is required to estimate conditional logit models \parencite{chamberlain1980analysis}. For all specifications, we find that the effects are precisely estimated, statistically significant and have the expected signs.
The estimates from the baseline specifications in columns (1) --- (3) indicate that a challenger having 10 additional units of reputation is 0.06 --- 0.14 percentage points more likely to persuade a poster on average, than another challenger of the same opinion with all else equal.
This corresponds to a 1.7 --- 4.0\% increase over the platform average debate success rate of 3.5\%.
Keeping in mind the difficulty of persuasion (the average persuasion rate of a user in our dataset is 0.6\%, and the median persuasion rate of a user is 0.0\%), this increase is significant.

Three additional observations are worth noting. First, we expect the estimated effect of skill to be attenuated due to measurement error in all specifications. Second, comparing columns (2) and (3), we find that including the opinion-fixed effects increases the estimated effect of reputation. This suggests that endogenous opinion selection (discussed in Section \ref{sec:baseline}) biases the estimated effect of reputation downwards. Second, the estimated effect of reputation is an average across all challengers, including those with low (and high) persuasive ability who are unlikely to benefit from additional reputation. We expect the impact of reputation to be higher for challengers with moderate persuasive ability.

The estimated effects from the baseline specifications in columns (1) --- (3) of Table \ref{tab:mainresults} may be confounded by unobserved time-varying challenger characteristics (discussed in Section \ref{sec:instrument}). Hence, we use the challenger's mean past position as an instrument for their reputation, and report the estimated local average treatment effects (LATEs) of reputation, skill and position on debate success with a linear instrumental variable specification in column (4).
The estimates indicate that a challenger having 10 additional units of reputation is 1.09 percentage points more likely on average to persuade a poster, than another challenger of the same opinion with all else equal. This corresponds to a 31\% increase over the platform average debate success rate of 3.5\%. Compared to the linear probability model estimates, the estimated LATE of reputation is 7.8 times larger. We attribute this increase to the compliers (debate subpopulations where the challenger's reputation is indeed affected by the instrument) having challengers with moderate to high persuasive ability, who benefit more from an increase in their reputation than those with low persuasive ability.

The instrument exclusion restriction will be violated if the mean response position of a challenger in the \textit{past} has a direct effect on their probability of success in the \textit{present} debate. This is possible if, for example, users become more persuasive by learning from the earlier responders to the opinions they challenged previously (as discussed in Section \ref{sec:instrument}). However, the persuasive ability acquired in this manner is likely to affect debate success through the text of the challenger's responses. Hence, we alleviate concerns of instrument exclusion being violated by controlling for the response text in a partially-linear instrumental variable specification, and report the estimated LATEs in column (5) of Table \ref{tab:mainresults}. The estimated LATEs from this specification are less precise and differ slightly from the estimates in column (4), but are statistically indistinguishable at the 5\% level (using a two-tailed Z-test). This suggests that the instrument exclusion restriction is not violated by factors present in the text.

We supplement these results by reporting LATE estimates using the plausibly-exogenous methodology \parencite{conley2012plausibly}, which assumes a known direct effect $\gamma$ of the instrument on debate success, for a range of $\gamma$ in Table \ref{tab:plausexogresults}.
If users become more persuasive by learning from earlier challengers, we expect $\gamma > 0$. Table \ref{tab:plausexogresults} indicates that the estimated reputation LATE after correcting for exclusion restriction violations of this type is \textit{larger}, in which case our linear instrumental variable specification (without this correction) under-estimates the true LATE. For $\gamma=-0.0002$,
the estimated LATE of reputation after correcting for the exclusion restriction violation becomes insignificant. This is expected, since correcting for $\gamma=-0.0002$ completely eliminates the ``reduced form'' effect of the instrument on debate success (reported in Appendix D). Setting the ``reduced form'' effect of the instrument on the outcome to zero as such necessarily renders the estimated treatment effect insignificant \parencite{angrist2001instrumental}.
For $\gamma < -0.0002$, the estimated LATE of reputation after correcting for exclusion restriction violations is negative, which is intuitively implausible.

In summary, our main results confirm hypothesis H1 (Section \ref{sec:conceptual}): that reputation has persuasive power on the ChangeMyView platform. In the rest of this section, we investigate potential mechanisms that could explain the persuasive power of reputation.

\clearpage

\addtolength{\tabcolsep}{-0.4mm}
\begin{table*}[!t]
\small
\centering
\caption{\textbf{Results with a plausibly-exogenous instrument.} Estimated effects of reputation, skill and position with a linear instrumental variable specification and plausibly-exogenous instrument. All specifications include opinion fixed-effects. Position is standardized to have zero-mean and unit-variance. The instrument $Z_{pu}$ is the mean past position of user $u$ before they challenged opinion $p$.}
\begin{tabular}[t]{lcccccc}
\toprule
& \multicolumn{6}{c}{Dependent Variable: Debate Success $Y_{pu}$}\\
\midrule
Exclusion violation ($\gamma$)& $-0.0004$$\phantom{^{***}}$ & $-0.0002$$\phantom{^{***}}$ & $-0.0001$$\phantom{^{***}}$ & $+0.0001$$\phantom{^{***}}$ & $+0.0002$$\phantom{^{***}}$ & $+0.0004$$\phantom{^{***}}$ \\
\midrule

Reputation $r_{pu}$ (10 units)
& $-0.0110$$^{***}$           & $-0.0001$$\phantom{^{***}}$           & $\phantom{-}0.0054$$^{***}$
& $\phantom{-}0.0163$$^{***}$ & $\phantom{-}0.0218$$^{***}$ & $\phantom{-}0.0327$$^{***}$ \\
& ($0.0007$)$\phantom{^*}$ & ($0.0006$)$\phantom{^*}$ & ($0.0006$)$\phantom{^*}$
& ($0.0006$)$\phantom{^*}$ & ($0.0007$)$\phantom{^*}$ & ($0.0008$)$\phantom{^*}$ \\[2mm]

Skill $s_{pu}$ (percentage)
& $\phantom{-}0.0062$$^{***}$           & $\phantom{-}0.0037$$^{***}$ & $\phantom{-}0.0025$$^{***}$
& $-0.0001$$\phantom{^{***}}$ & $-0.0013$$^{***}$           & $-0.0038$$^{***}$ \\
& ($0.0002$)$\phantom{^*}$ & ($0.0002$)$\phantom{^*}$ & ($0.0002$)$\phantom{^*}$
& ($0.0002$)$\phantom{^*}$ & ($0.0002$)$\phantom{^*}$ & ($0.0002$)$\phantom{^*}$ \\[2mm]

Position $t_{pu}$ (std. deviations)
& $-0.0116$$^{***}$ & $-0.0097$$^{***}$ & $-0.0087$$^{***}$
& $-0.0069$$^{***}$ & $-0.0059$$^{***}$ & $-0.0040$$^{***}$ \\
& ($0.0010$)$\phantom{^*}$ & ($0.0009$)$\phantom{^*}$ & ($0.0008$)$\phantom{^*}$
& ($0.0007$)$\phantom{^*}$ & ($0.0007$)$\phantom{^*}$ & ($0.0006$)$\phantom{^*}$ \\[2mm]

Response text ($X_{pu}$)               & \xmark      & \xmark      & \xmark      & \xmark      & \xmark      & \xmark      \\[1mm]
Instrument ($Z_{pu}$)                  & \cmark      & \cmark      & \cmark      & \cmark      & \cmark      & \cmark      \\[1mm]
Opinion fixed-effects ($\tau_p$)       & \cmark      & \cmark      & \cmark      & \cmark      & \cmark      & \cmark      \\[1mm]
No. of debates                         & $1,109,469$ & $1,109,469$ & $1,109,469$ & $1,109,469$ & $1,109,469$ & $1,109,469$ \\
\bottomrule
\end{tabular}
\label{tab:plausexogresults}
\vspace{3mm}

Note: Standard errors (clustered by opinion) displayed in parentheses. $^{***}p<0.001; ^{**}p<0.01; ^{*}p<0.05$
\end{table*}%
\addtolength{\tabcolsep}{0.4mm}

Motivated by a theoretical model of persuasion with reference cues \parencite{bilancini2018rational}, we investigate whether reputation serves as a reference cue (an information-processing shortcut) for the challenger's response quality by examining patterns in the heterogeneity of the effects of reputation and skill with the content of the debate. These patterns reflect how the poster's usage of heuristic and systematic information-processing \parencite{chaiken1989heuristic} varies with content characteristics that affect information-processing effort.
The model predicts that individuals rely more on low-effort heuristic processing than on high-effort systematic processing when subject to greater cognitive overload. In our setting, this translates to posters relying more on a challenger's reputation than on their skill when the challenger's response is more cognitively complex (hypothesis H2 in Section \ref{sec:conceptual}). The model also predicts that individuals rely less on heuristic processing than on systematic processing when they are more involved in the issue being debated (hypothesis H3 in Section \ref{sec:conceptual}).

We test both predictions by examining how the relative estimated effects of reputation and skill on debate success vary with the cognitive complexity of the challenger's response and with the issue-involvement of the poster. Using the length (in characters) of the challenger's response text and of the poster's opinion text as proxies for cognitive complexity and issue-involvement, respectively, Table \ref{tab:heterogeneity} reports LATE estimates of the effects of reputation and skill interacted with the response and opinion length (binned into quantiles). Note that the specifications in Table \ref{tab:heterogeneity} do not include opinion fixed-effects, which would absorb all variation in the opinion length. They include calendar month-year fixed-effects, since unobserved temporal variation is no longer accounted for without the opinion fixed-effects.

\clearpage

The estimated LATEs in columns (1) and (2) of Table \ref{tab:heterogeneity} quantify how the effects of reputation and skill (separately) vary with the response length, which serves as a proxy for the cognitive complexity of the challenger's response. The effects of both reputation and skill increase with the response length. This is expected since, with longer responses, the content explains more of the debate outcome than other factors. However, the effect of reputation increases more than that of skill from the first to the fourth response length quantile: by +0.0133 for reputation compared to +0.0030 for skill. Measuring the share of the effect magnitude of reputation (of the sum of effect magnitudes of reputation and skill) at each response length quantile reveals that the reputation effect share increases from 82\% to 89\%. This is consistent with the poster's increased reliance on reputation as a heuristic shortcut and decreased reliance on systematic evaluation of argument quality. This pattern supports hypothesis H2.

The estimates in columns (3) and (4) quantify how the effects of reputation and skill vary with the opinion length, where we assume that longer opinions are correlated with the poster being more involved in the issue being debated. The effect of reputation remains largely the same after the second opinion length quantile, while that of of skill increases significantly in each subsequent opinion length quantile. The share of the effect magnitude of reputation (of the sum of effect magnitudes of reputation and skill) decreases from 90\% to 83\% from the second to the fourth opinion length quantile. This is consistent with the decreased reliance of the poster on reputation as a heuristic shortcut and the increased reliance on systematic evaluation of argument quality as the issue-involvement of the poster increases. This pattern supports hypothesis H3.

Similar trends are observed if reading complexity measures (such as the Flesch-Kincaid Reading Ease) are used to proxy for the cognitive complexity of the challenger's response and the issue-involvement of the poster. The negative estimated effects of skill for opinions and responses in the first length quantile could be attributed to higher skilled users preferring to write longer responses and to challenge longer opinions. These preferences are a form of endogenous selection on the interaction terms, and will bias our estimates downwards. Since we only examine trends in the estimated effects, and do not interpret their absolute values, these biases are not a major concern.

We now examine if reputation serves as a way for cognitively-overloaded posters to select challengers to engage with.
Table \ref{tab:ocheterogeneity} reports the LATE estimates of reputation interacted with the total number of challengers of the opinion, binned into quantiles. The specification in Table \ref{tab:ocheterogeneity} does not include opinion fixed-effects, since they absorb all variation in the number of challengers, but includes calendar month-year fixed-effects and the response and opinion length as controls. We expect that, under the larger cognitive burden of having to respond to more challengers, posters would rely on reputation as a filtering heuristic. However, the estimates in Table \ref{tab:ocheterogeneity} indicate a decrease in the effect of reputation as the number of opinion challengers increases, which likely reflects the preference of reputed users for challenging opinions with fewer existing challengers. Hence, we find no support for the hypothesis that posters use reputation to select challengers to engage with.

\clearpage
\thispagestyle{empty}

\addtolength{\tabcolsep}{3.25mm}
\begin{table*}[!t]
\vspace{-8mm}
\small
\centering
\caption{\textbf{Heterogeneity with response and opinion length.} Variation in the estimated LATEs of reputation, skill and position on debate success with  response and opinion length (binned into quantiles).}
\begin{tabular}[t]{lcccc}
\toprule
& \multicolumn{4}{c}{Dependent Variable: Debate Success $Y_{pu}$}\\
\midrule
Moderator $M_{pu}$ & \multicolumn{2}{c}{Response Text Length} & \multicolumn{2}{c}{Opinion Text Length} \\
\midrule
& (1) & (2) & (3) & (4) \\
\midrule
Reputation $r_{pu}$ (10 units)
&
& $\phantom{-}0.0093$$^{***}$
&
& $\phantom{-}0.0085$$^{***}$\\
&
& ($0.0005$)$\phantom{^{*}}$
&
& ($0.0005$)$\phantom{^{*}}$\\[0mm]

  \qquad $\times ~ \mathbb{I}[M_{pu} \in 1^{\textrm{st}} \textrm{ quantile}]$
  & $\phantom{-}0.0038$$^{***}$
  &
  & $-0.0007$$\phantom{^{***}}$
  & \\
  & ($0.0006$)$\phantom{^{*}}$
  &
  & ($0.0006$)$\phantom{^{*}}$
  & \\[0mm]

  \qquad $\times ~ \mathbb{I}[M_{pu} \in 2^{\textrm{nd}} \textrm{ quantile}]$
  & $\phantom{-}0.0066$$^{***}$
  &
  & $\phantom{-}0.0114$$^{***}$
  & \\
  & ($0.0006$)$\phantom{^{*}}$
  &
  & ($0.0007$)$\phantom{^{*}}$
  & \\[0mm]

  \qquad $\times ~ \mathbb{I}[M_{pu} \in 3^{\textrm{rd}} \textrm{ quantile}]$
  & $\phantom{-}0.0088$$^{***}$
  &
  & $\phantom{-}0.0117$$^{***}$
  & \\
  & ($0.0006$)$\phantom{^{*}}$
  &
  & ($0.0007$)$\phantom{^{*}}$
  & \\[0mm]

  \qquad $\times ~ \mathbb{I}[M_{pu} \in 4^{\textrm{th}} \textrm{ quantile}]$
  & $\phantom{-}0.0171$$^{***}$
  &
  & $\phantom{-}0.0126$$^{***}$
  & \\
  & ($0.0010$)$\phantom{^{*}}$
  &
  & ($0.0008$)$\phantom{^{*}}$
  & \\[1mm]

Skill $s_{pu}$ (percentage)
& $\phantom{-}0.0005$$^{***}$
&
& $\phantom{-}0.0009$$^{***}$
& \\
& ($0.0002$)$\phantom{^{*}}$
&
& ($0.0001$)$\phantom{^{*}}$
& \\[0mm]

  \qquad $\times ~ \mathbb{I}[M_{pu} \in 1^{\textrm{st}} \textrm{ quantile}]$
  &
  & $-0.0008$$^{***}$
  &
  & $-0.0021$$^{***}$
  \\
  &
  & ($0.0002$)$\phantom{^{*}}$
  &
  & ($0.0002$)$\phantom{^{*}}$
  \\[0mm]

  \qquad $\times ~ \mathbb{I}[M_{pu} \in 2^{\textrm{nd}} \textrm{ quantile}]$
  &
  & $-0.0002$$\phantom{^{***}}$
  &
  & $\phantom{-}0.0012$$^{***}$
  \\
  &
  & ($0.0002$)$\phantom{^{*}}$
  &
  & ($0.0002$)$\phantom{^{*}}$
  \\[0mm]

  \qquad $\times ~ \mathbb{I}[M_{pu} \in 3^{\textrm{rd}} \textrm{ quantile}]$
  &
  & $\phantom{-}0.0001$$\phantom{^{***}}$
  &
  & $\phantom{-}0.0018$$^{***}$
  \\
  &
  & ($0.0002$)$\phantom{^{*}}$
  &
  & ($0.0002$)$\phantom{^{*}}$
  \\[0mm]

  \qquad $\times ~ \mathbb{I}[M_{pu} \in 4^{\textrm{th}} \textrm{ quantile}]$
  &
  & $\phantom{-}0.0022$$^{***}$
  &
  & $\phantom{-}0.0025$$^{***}$
  \\
  &
  & ($0.0002$)$\phantom{^{*}}$
  &
  & ($0.0002$)$\phantom{^{*}}$
  \\[1mm]

Position $t_{pu}$ (std. deviations)
& $-0.0074$$^{***}$
& $-0.0071$$^{***}$
& $-0.0071$$^{***}$
& $-0.0071$$^{***}$ \\
& ($0.0008$)$\phantom{^{*}}$
& ($0.0008$)$\phantom{^{*}}$
& ($0.0008$)$\phantom{^{*}}$
& ($0.0008$)$\phantom{^{*}}$ \\[1mm]

Response Length (characters)
&
&
&
&
\\[2mm]
  \qquad $\mathbb{I}[\in 1^{\textrm{st}} \textrm{ quantile}]$
  &
  &
  &
  &
  \\
  &
  &
  &
  &
  \\[0mm]

  \qquad $\mathbb{I}[\in 2^{\textrm{nd}} \textrm{ quantile}]$
  & $\phantom{-}0.0035$$^{***}$
  & $\phantom{-}0.0046$$^{***}$
  & $\phantom{-}0.0072$$^{***}$
  & $\phantom{-}0.0060$$^{***}$
  \\
  & ($0.0009$)$\phantom{^{*}}$
  & ($0.0005$)$\phantom{^{*}}$
  & ($0.0004$)$\phantom{^{*}}$
  & ($0.0004$)$\phantom{^{*}}$
  \\[0mm]

  \qquad $\mathbb{I}[\in 3^{\textrm{rd}} \textrm{ quantile}]$
  & $\phantom{-}0.0080$$^{***}$
  & $\phantom{-}0.0119$$^{***}$
  & $\phantom{-}0.0157$$^{***}$
  & $\phantom{-}0.0143$$^{***}$
  \\
  & ($0.0010$)$\phantom{^{*}}$
  & ($0.0006$)$\phantom{^{*}}$
  & ($0.0005$)$\phantom{^{*}}$
  & ($0.0005$)$\phantom{^{*}}$
  \\[0mm]

  \qquad $\mathbb{I}[\in 4^{\textrm{th}} \textrm{ quantile}]$
  & $\phantom{-}0.0218$$^{***}$
  & $\phantom{-}0.0306$$^{***}$
  & $\phantom{-}0.0395$$^{***}$
  & $\phantom{-}0.0396$$^{***}$
  \\
  & ($0.0013$)$\phantom{^{*}}$
  & ($0.0007$)$\phantom{^{*}}$
  & ($0.0006$)$\phantom{^{*}}$
  & ($0.0006$)$\phantom{^{*}}$
  \\[0mm]

  Opinion Length (characters)
  &
  &
  &
  &\\[2mm]

    \qquad $\mathbb{I}[\in 1^{\textrm{st}} \textrm{ quantile}]$
    &
    &
    &
    &
    \\
    &
    &
    &
    &
    \\[0mm]

    \qquad $\mathbb{I}[\in 2^{\textrm{nd}} \textrm{ quantile}]$
    & $\phantom{-}0.0257$$^{***}$
    & $\phantom{-}0.0262$$^{***}$
    & $\phantom{-}0.0036$$^{*}\phantom{^{**}}$
    & $\phantom{-}0.0153$$^{***}$
    \\
    & ($0.0008$)$\phantom{^{*}}$
    & ($0.0008$)$\phantom{^{*}}$
    & ($0.0014$)$\phantom{^{*}}$
    & ($0.0008$)$\phantom{^{*}}$
    \\[0mm]

    \qquad $\mathbb{I}[\in 3^{\textrm{rd}} \textrm{ quantile}]$
    & $\phantom{-}0.0284$$^{***}$
    & $\phantom{-}0.0289$$^{***}$
    & $\phantom{-}0.0054$$^{***}$
    & $\phantom{-}0.0161$$^{***}$
    \\
    & ($0.0008$)$\phantom{^{*}}$
    & ($0.0008$)$\phantom{^{*}}$
    & ($0.0014$)$\phantom{^{*}}$
    & ($0.0008$)$\phantom{^{*}}$
    \\[0mm]

    \qquad $\mathbb{I}[\in 4^{\textrm{th}} \textrm{ quantile}]$
    & $\phantom{-}0.0309$$^{***}$
    & $\phantom{-}0.0316$$^{***}$
    & $\phantom{-}0.0068$$^{***}$
    & $\phantom{-}0.0167$$^{***}$
    \\
    & ($0.0008$)$\phantom{^{*}}$
    & ($0.0008$)$\phantom{^{*}}$
    & ($0.0016$)$\phantom{^{*}}$
    & ($0.0008$)$\phantom{^{*}}$
    \\[1mm]

Response text ($X_{pu}$)                   & \xmark      & \xmark      & \xmark      & \xmark      \\[0.5mm]
Instrument ($Z_{pu}$)                      & \cmark      & \cmark      & \cmark      & \cmark      \\[0.5mm]
Opinion fixed-effects ($\tau_p$)           & \xmark      & \xmark      & \xmark      & \xmark      \\[0.5mm]
Month-year fixed-effects                   & \cmark      & \cmark      & \cmark      & \cmark      \\[0.5mm]
No. of debates                             & $1,026,201$ & $1,026,201$ & $1,026,201$ & $1,026,201$ \\
\bottomrule
\end{tabular}
\label{tab:heterogeneity}
\vspace{3mm}

Note: Standard errors (clustered by opinion) displayed in parentheses. $^{***}p<0.001; ^{**}p<0.01; ^{*}p<0.05$
\vspace{0mm}
\end{table*}%
\addtolength{\tabcolsep}{-3.25mm}

\clearpage
\thispagestyle{empty}

\addtolength{\tabcolsep}{2.25mm}
\begin{table*}[!t]
\vspace{-8mm}
\small
\centering
\caption{\textbf{Heterogeneity with the number of opinion challengers.} Variation in the estimated LATEs of reputation, skill and position with the number of opinion challengers (binned into quantiles).}
\begin{tabular}[t]{lc}
\toprule
& Dependent Variable: Debate Success $Y_{pu}$\\
\midrule
Reputation $r_{pu}$ (10 units) & \\[2mm]
  \qquad $\times ~ \mathbb{I}[\textrm{Number of opinion challengers } \in 1^{\textrm{st}} \textrm{ quantile}]$
  & $\phantom{-}0.0086$$^{***}$ \\
  & ($0.0008$)$\phantom{^{*}}$ \\[0mm]

  \qquad $\times ~ \mathbb{I}[\textrm{Number of opinion challengers } \in 2^{\textrm{nd}} \textrm{ quantile}]$
  & $\phantom{-}0.0082$$^{***}$ \\
  & ($0.0007$)$\phantom{^{*}}$ \\[0mm]

  \qquad $\times ~ \mathbb{I}[\textrm{Number of opinion challengers } \in 3^{\textrm{rd}} \textrm{ quantile}]$
  & $\phantom{-}0.0067$$^{***}$ \\
  & ($0.0006$)$\phantom{^{*}}$ \\[0mm]

  \qquad $\times ~ \mathbb{I}[\textrm{Number of opinion challengers } \in 4^{\textrm{th}} \textrm{ quantile}]$
  & $\phantom{-}0.0044$$^{***}$ \\
  & ($0.0012$)$\phantom{^{*}}$ \\[2mm]

Skill $s_{pu}$ (percentage)
& $\phantom{-}0.0009$$^{***}$ \\
& ($0.0001$)$\phantom{^{*}}$ \\[2mm]

Position $t_{pu}$ (std. deviations)
& $-0.0055$$^{***}$ \\
& ($0.0008$)$\phantom{^{*}}$ \\[0mm]

Number of opinion-challengers
& \\[2mm]
  \qquad $\mathbb{I}[\in 1^{\textrm{st}} \textrm{ quantile}]$
  & \\
  & \\[0mm]

  \qquad $\mathbb{I}[\in 2^{\textrm{nd}} \textrm{ quantile}]$
  & $-0.0101$$^{***}$ \\
  & ($0.0021$)$\phantom{^{*}}$  \\[0mm]

  \qquad $\mathbb{I}[\in 3^{\textrm{rd}} \textrm{ quantile}]$
  & $-0.0178$$^{***}$ \\
  & ($0.0019$)$\phantom{^{*}}$  \\[0mm]

  \qquad $\mathbb{I}[\in 4^{\textrm{th}} \textrm{ quantile}]$
  & $-0.0144$$^{***}$ \\
  & ($0.0022$)$\phantom{^{*}}$  \\[0mm]

Response Length (characters)
  & \\[2mm]
    \qquad $\mathbb{I}[\in 1^{\textrm{st}} \textrm{ quantile}]$
    & \\
    & \\[0mm]

    \qquad $\mathbb{I}[\in 2^{\textrm{nd}} \textrm{ quantile}]$
    & $\phantom{-}0.0061$$^{***}$ \\
    & ($0.0004$)$\phantom{^{*}}$  \\[0mm]

    \qquad $\mathbb{I}[\in 3^{\textrm{rd}} \textrm{ quantile}]$
    & $\phantom{-}0.0145$$^{***}$ \\
    & ($0.0005$)$\phantom{^{*}}$  \\[0mm]

    \qquad $\mathbb{I}[\in 4^{\textrm{th}} \textrm{ quantile}]$
    & $\phantom{-}0.0401$$^{***}$ \\
    & ($0.0006$)$\phantom{^{*}}$  \\[0mm]

Opinion Length (characters)
  & \\[2mm]
    \qquad $\mathbb{I}[\in 1^{\textrm{st}} \textrm{ quantile}]$
    & \\
    & \\[0mm]

    \qquad $\mathbb{I}[\in 2^{\textrm{nd}} \textrm{ quantile}]$
    & $\phantom{-}0.0271$$^{***}$ \\
    & ($0.0008$)$\phantom{^{*}}$  \\[0mm]

    \qquad $\mathbb{I}[\in 3^{\textrm{rd}} \textrm{ quantile}]$
    & $\phantom{-}0.0298$$^{***}$ \\
    & ($0.0008$)$\phantom{^{*}}$  \\[0mm]

    \qquad $\mathbb{I}[\in 4^{\textrm{th}} \textrm{ quantile}]$
    & $\phantom{-}0.0324$$^{***}$ \\
    & ($0.0008$)$\phantom{^{*}}$  \\[2mm]

Response text ($X_{pu}$)                   & \xmark      \\[1mm]
Instrument ($Z_{pu}$)                      & \cmark      \\[1mm]
Opinion fixed-effects ($\tau_p$)           & \xmark      \\[1mm]
Month-year fixed-effects                   & \cmark      \\[1mm]
No. of debates                             & $1,026,201$ \\
\bottomrule
\end{tabular}
\label{tab:ocheterogeneity}
\vspace{3mm}

Note: Standard errors (clustered by opinion) displayed in parentheses. $^{***}p<0.001; ^{**}p<0.01; ^{*}p<0.05$
\vspace{0mm}
\end{table*}%
\addtolength{\tabcolsep}{-2.25mm}

\clearpage

\addtolength{\tabcolsep}{5.5mm}
\begin{table*}[!h]
\small
\centering
\caption{\textbf{Conversation tree length as the outcome.} Estimated effects of reputation, skill and position on the conversation tree length with the linear instrumental variable specification. All specifications include opinion fixed-effects. Position is standardized to have zero-mean and unit-variance. The instrument $Z_{pu}$ is the mean past position of user $u$ before they challenged opinion $p$.}
\begin{tabular}[t]{lccc}
\toprule
& \multicolumn{3}{c}{Dependent Variable: Conversation Tree Length}\\
\midrule
& (1) & (2) & (3) \\\midrule
Reputation $r_{pu}$ (10 units)
& $\phantom{-}0.8537$$^{***}$
& $\phantom{-}0.8499$$^{***}$
& $\phantom{-}0.3661$$^{**}\phantom{^*}$ \\
& ($0.025$)
& ($0.025$)
& ($0.138$) \\[2mm]

Skill $s_{pu}$ (percentage)
& $-0.1123$$^{***}$
& $-0.1184$$^{***}$
& $-0.0273$$\phantom{^{***}}$ \\
& ($0.005$)
& ($0.005$)
& ($0.027$) \\[2mm]

Position $t_{pu}$ (std. deviations)
& $-1.2354$$^{***}$
& $-1.1888$$^{***}$
& $-2.6737$$^{***}$ \\
& ($0.103$)
& ($0.100$)
& ($0.511$) \\[2mm]

Conditional on debate success $Y_{pu}$     &&&\\[1mm]
\qquad Unsuccessful debates ($Y_{pu} = 0$) & \xmark      & \cmark      & \xmark      \\[1mm]
\qquad Successful debates ($Y_{pu} = 1$)   & \xmark      & \xmark      & \cmark      \\[1mm]
Response text ($X_{pu}$)                   & \xmark      & \xmark      & \xmark      \\[1mm]
Instrument ($Z_{pu}$)                      & \cmark      & \cmark      & \cmark      \\[1mm]
Opinion fixed-effects ($\tau_p$)           & \cmark      & \cmark      & \cmark      \\[1mm]
No. of debates                             & $1,109,469$ & $982,867$ & $23,157$ \\
\bottomrule
\end{tabular}
\label{tab:altoutcomes}
\vspace{3mm}

Note: Standard errors (clustered by opinion) displayed in parentheses. $^{***}p<0.001; ^{**}p<0.01; ^{*}p<0.05$
\vspace{0mm}
\end{table*}%
\addtolength{\tabcolsep}{-5.5mm}
We also examine if having higher reputation leads to longer conversations with the challenger, which could mediate the effect of reputation on debate success. We estimate the effect of reputation on the \textit{conversation tree length}: the maximum number of turns of dialogue in the conversation tree
initiated by the challenger's response.
It equals 1 if no one responds to the challenger, and is greater than 1 otherwise. It is positively (but weakly) correlated with debate success ($r=0.13, p<0.001$).

Table \ref{tab:altoutcomes} reports the estimated LATEs of reputation, skill and position on the conversation tree length.
The estimates in column (1) indicate that having 10 additional units of reputation leads to conversations trees that are 0.85 turns longer on average,with all else equal. Since the poster must use one turn of dialogue when awarding a $\Delta$ to the challenger, this estimate may simply reflect the direct effect of reputation on persuading the poster. Hence, columns (2) and (3) report the estimated LATEs separately for unsuccessful and successful debates,
and find that having 10 additional units of reputation leads to conversations trees that are 0.85 turns longer on average even for unsuccessful debates.
Hence, it is plausible that reputation affects debate success via longer conversations.

The estimated LATEs also indicate that an additional percentage point of skill leads to conversation trees that are up to 0.11 turns shorter on average for both unsuccessful and successful debates. This suggests that higher skilled challengers are either quicker to abandon futile conversations, or are able to persuade the poster in fewer turns of dialogue.

\clearpage

\addtolength{\tabcolsep}{6mm}
\begin{table*}[!t]
\small
\centering
\caption{\textbf{Single-party vs. multi-party debates.} Estimated effects of reputation, skill and position on whether a debate is multi-party in column (1), and on debate success for single-party debates only in column (2), with a linear instrumental variable specification. All specifications include opinion fixed-effects. Position is standardized to have zero-mean and unit-variance. The instrument $Z_{pu}$ is the mean past position of user $u$ before they challenged opinion $p$.}
\begin{tabular}[t]{lcc}
\toprule
Dependent Variable & $\mathbb{I}[$ Debate $pu$ is multi-party $]$ & Debate Success $Y_{pu}$\\
\midrule
                   & (1) & (2)\\
\midrule
Reputation $r_{pu}$ (10 units)
& $\phantom{-}0.1093$$^{***}$
& $\phantom{-}0.0062$$^{***}$\\
& ($0.0034$)$\phantom{^{*}}$
& ($0.0006$)$\phantom{^{*}}$\\[2mm]

Skill $s_{pu}$ (percentage)
& $-0.0167$$^{***}$
& $\phantom{-}0.0010$$^{***}$\\
& ($0.0007$)$\phantom{^{*}}$
& ($0.0002$)$\phantom{^{*}}$\\[2mm]

Position $t_{pu}$ (std. deviations)
& $-0.1781$$^{***}$
& $-0.0021$$^{***}$\\
& ($0.0148$)$\phantom{^{*}}$
& ($0.0003$)$\phantom{^{*}}$\\[2mm]

Only single-party debates                  & \xmark      & \cmark    \\[1mm]
Response text ($X_{pu}$)                   & \xmark      & \xmark    \\[1mm]
Instrument ($Z_{pu}$)                      & \cmark      & \cmark    \\[1mm]
Opinion fixed-effects ($\tau_p$)           & \cmark      & \cmark    \\[1mm]
No. of debates                             & $1,109,468$ & $667,678$ \\
\bottomrule
\end{tabular}
\label{tab:multiparty}
\vspace{3mm}

Note: Standard errors (clustered by opinion) displayed in parentheses. $^{***}p<0.001; ^{**}p<0.01; ^{*}p<0.05$
\vspace{0mm}
\end{table*}%
\addtolength{\tabcolsep}{-6mm}

Finally, we examine the effect of reputation on attracting collaboration from other users. Recall from Section \ref{sec:data} that other (non-poster) users may join an ongoing debate between the challenger and poster; we term such debates \textit{multi-party}. 6.1\% of multi-party debates are successful, as compared to 2.4\% of non-multi-party debates, indicating a positive association between debate success and whether the debate is multi-party. If this association is causal, part of the effect of reputation on debate success may be due to attracting collaboration from other users. The estimated LATEs in column (1) of Table \ref{tab:multiparty} indicate that having 10 additional units of reputation increases the probability of the debate being multi-party by 11\%. The mechanism via which higher reputation challengers attract other users to join their debates may be complex. For example, higher reputation challengers may engage in longer conversations that make it easier for other users to join.

To exclude the effect of reputation that is potentially
due to attracting collaboration from other users, we report the estimated LATEs of reputation, skill and position on debate success for single-party debates only in column (2) of Table \ref{tab:multiparty}. The estimates share the same sign as the overall LATEs reported in Table \ref{tab:mainresults}, with smaller magnitudes. This could be due to single-party debates occurring on less popular topics, which are more difficult to persuade in, and hence less likely to attract users who may collaborate with the challengers in ongoing debates. Nevertheless, reputation continues to have a significant positive effect on debate success after excluding the potential effect of collaboration from other non-poster users.

\section{Conclusion}
\label{sec:conclusion}

Using a 7-year panel of over a million debates from a large online argumentation platform containing explicit indicators of successful persuasion, we quantify the persuasive power of \textit{ethos} --- one of the three modes of persuasion described in Aristotle's \textit{Rhetoric} --- in deliberation online.
Specifically, we identify the causal effect of the reputation of an opinion challenger on persuading a poster on the ChangeMyView platform, using the mean position of the challenger in past debates as an instrument for their present reputation. We address endogenous opinion selection using opinion fixed-effects, and ensure instrument validity by controlling for the text of the challenger's response using neural models of text in a partially-linear instrumental variable specification.

We find a statistically significant positive effect of reputation on debate success, with 10 additional units of reputation increasing the probability of persuading the poster by 31\% over the mean platform persuasion rate of 3.5\%. The relative effect of the challenger's reputation with respect to their skill increases with the cognitive complexity of their response (proxied for by the length of their response text), and decreases with the issue-involvement of the poster (proxied for by the length of the opinion text), confirming the predictions of a theoretical model of persuasion where reputation serves as a reference cue \parencite{bilancini2018rational}.
We find no evidence that posters use reputation as a way to select which challengers to engage with, but we do find evidence that reputation induces longer conversations with the challenger. While we find evidence that reputation attracts collaboration with the challenger from other users (which may in turn affect persuasion), the estimated effect of reputation continues to be positive and statistically significant after excluding debates where such collaboration occurred. These findings suggest that reputation serves as a proxy for argument quality and validity, and is used by cognitively-overloaded posters as a low-effort information-processing heuristic to evaluate a challenger's arguments.

Our findings have implications for a variety of platforms that facilitate deliberative decision-making online,
including the Stanford Online Deliberation Platform\footnote{\url{https://stanforddeliberate.org/}} used by government organizations to elicit public opinion, and Github used by technology firms engaged in remote collaborative software development \parencite{marlow2013impression}. Specifically, our results suggest that if the participants in deliberation sessions can observe characteristics of other participants, such as past organizational contributions or awards, such characteristics may serve as reference-cues that endow ``reputed'' participants with additional persuasive power.
As a consequence, the degree of consensus within each deliberation session could increase, while distorting the average consensus opinions over a sequence of deliberation sessions towards those held by reputed participants. Such an outcome is undesirable in practice, and violates one of the key tenets of authentic deliberation \parencite{fishkin2005experimenting}. In general, our findings may be of interest to online platforms that employ reputation systems and are concerned about their unintended effects \parencite{shi2020reputation,chen2018user}.

\clearpage

The reference-cue mechanism \parencite{bilancini2018rational} supported by our findings reveals three possible managerial strategies to address this inequity. Importantly, these strategies do not rely on simply hiding all reputation indicators, since employees may desire to showcase their reputation for several justifiable reasons without having any intention to exploit its persuasive power. Visible indicators of reputation are also potent motivators, and hiding them completely could nullify their ability to incentivize engagement. Instead, organizations can increase the effort required to view participants' reputation by displaying it less prominently, albeit at the cost of reducing the motivational power of reputation. Alternatively, firms can simplify the language of participants' arguments to reduce their information-processing costs (with machine learning tools like Grammarly, for example), thus reducing the need for participants to rely on low-effort heuristics. This, however, comes at the cost of disrupting participants' natural communication styles. Finally, firms can manipulate the perceived accuracy of reputation as an information-processing shortcut by using ambiguous reputation displays (using colors instead of numbers, for example). In contrast with the previous strategies, this strategy is simple to implement and unlikely to have negative repercussions.

Our findings also have implications for organizations directly engaged in online persuasion via dialogue, for purposes as varied as sales, addressing customer complaints \parencite{hauser1983defensive,ma2015squeaky}, and reducing societal polarization.
Since individuals online are often under information-overload, they are likely to rely on heuristic signals when faced with a persuasive message. Hence, low-reputation organizations (such as new entrants in a market, or organizations having a less favorable reputation among certain demographics) could benefit from simplifying their messages to induce less cognitive overload. Alternatively, organizations could benefit from acquiring reputation signals that are perceived as a proxy for their quality. However, it is important to determine which reputation signals would be perceived as accurate proxies by the organizations' target demographic and require relatively low effort to evaluate.

The ChangeMyView argumentation platform we study provides us a rich source of online conversations containing explicit indicators of persuasion. The large scale, long timeframe and high quality of the textual data enable identifying and examining causal mechanisms that would be impossible outside of laboratory settings. However, this comes at the cost of generalizing to other online (and offline) settings. The heavily moderated, good-faith discussions on this platform are different from the typical conversations on social media platforms like Facebook and Twitter, that are dominated by hostility and groupthink. Hence, our estimates could be viewed as upper bounds on the effects of reputation and skill on persuasion, and should be validated by future research into situations were successful persuasion is considered near-impossible or completely independent of the message content. Likewise, additional investigation is needed to study the effect of reputation on conversations offline, where a host of other factors (such as physical appearances and the environmental setting) are likely to affect successful persuasion.

\clearpage

\printbibliography

\clearpage

\section*{Appendix A: Platform Rules}

\noindent \textbf{Rules for shared opinions:}
\begin{itemize}
  \item \textbf{Rule A:} Explain the reasoning behind your view (500+ characters required).
  \item \textbf{Rule B:} You must personally hold the view and demonstrate that you are open to it changing.
  \item \textbf{Rule C:} Submission titles must adequately sum up your view.
  \item \textbf{Rule D:} Posts cannot express a neutral stance, suggest harm against a specific person, or be self-promotional.
  \item \textbf{Rule E:} Only post if you are willing to have a conversation with those who reply to you, and are available to do so within 3 hours after posting.
\end{itemize}

\noindent An opinion may also be removed if it violates any of the following:
\begin{enumerate}
  \item It was posted by a brand new account on a highly controversial topic.
  \item The user already has an active opinion from the last 24 hours. This is to encourage posters to stay engaged with their posts and continue discussion.
  \item It's identical in principle to another post made within 24 hours before it.
  \item Anything that is clearly spam or posted by a bot/novelty account.
\end{enumerate}

\noindent Opinions will never be removed based on topic or content, so long as they follow the rules above.

\noindent \textbf{Rules for responses to opinions:}
\begin{itemize}
  \item \textbf{Rule 1:} Direct responses to a submission must challenge or question at least one aspect of the submitted view.
  \item \textbf{Rule 2:} Don't be rude or hostile to other users.
  \item \textbf{Rule 3:} Refrain from accusing the poster or anyone else of being unwilling to change their view.
  \item \textbf{Rule 4:} Award a delta when acknowledging a change in your view, and not for any other reason.
  \item \textbf{Rule 5:} Responses must contribute meaningfully to the conversation.
\end{itemize}

\noindent A response may also be removed if it violates any of the following:
\begin{enumerate}
   \item It is a deliberate attempt to disrupt discussion.
   \item Anything that is clearly spam or posted by a bot/novelty account.
\end{enumerate}

\clearpage

\section*{Appendix B: Correlation Table}

All correlations are significant at $p<0.001$.

\addtolength{\tabcolsep}{3.0mm}
\begin{table*}[!h]
\small
\centering
\begin{tabular}[p]{lccccc}
\toprule
& (1) & (2) & (3) & (4) & (5) \\
\midrule
(1) Reputation                    & \phantom{-}1.00 & ---             & ---             & ---             & ---              \\
(2) Skill                         & \phantom{-}0.29 & \phantom{-}1.00 & ---             & ---             & ---              \\
(3) Position                      & -0.11           & -0.14           & \phantom{-}1.00 & ---             & ---              \\
(4) Mean past position            & -0.11           & -0.14           & \phantom{-}0.22 & \phantom{-}1.00 & ---              \\
(5) Number of past debates        & \phantom{-}0.88 & \phantom{-}0.16 & -0.12           & -0.10           & \phantom{-}1.00  \\
\bottomrule
\end{tabular}
\end{table*}%
\addtolength{\tabcolsep}{-3.0mm}

\section*{Appendix C: Results Including Debates by Deleted Challengers}

\addtolength{\tabcolsep}{3mm}
\begin{table*}[!h]
\small
\centering
\caption{\textbf{Main results.} Estimated effects of reputation, skill and position on debate success with a logit model without opinion fixed-effects (1), linear probability models without (2) and with (3) opinion fixed-effects, linear instrumental variable (4) specifications. Position is standardized to have zero-mean and unit-variance. Average marginal effects and pseudo-$R^2$ are reported for the logit model. The instrument $Z_{pu}$ is the mean past position of user $u$ before they challenged opinion $p$.}
\begin{tabular}[t]{lcccc}
\toprule
& \multicolumn{4}{c}{Dependent Variable: Debate Success $Y_{pu}$}\\
\midrule
& (1) & (2) & (3) & (4) \\
\midrule

Reputation $r_{pu}$ (10 units)
& $\phantom{-}0.0007$$^{***}$
& $\phantom{-}0.0011$$^{***}$
& $\phantom{-}0.0015$$^{***}$
& $\phantom{-}0.0068$$^{***}$ \\
& ($0.00003$) & ($0.00006$)
& ($0.00006$) & ($0.00081$) \\[2mm]

Skill $s_{pu}$ (percentage)
& $\phantom{-}0.0024$$^{***}$
& $\phantom{-}0.0042$$^{***}$
& $\phantom{-}0.0036$$^{***}$
& $\phantom{-}0.0023$$^{***}$ \\
& ($0.00004$) & ($0.00007$)
& ($0.00007$) & ($0.00021$)\\[2mm]

Position $t_{pu}$ (std. deviations)
& $-0.0446$$^{***}$ & $-0.0085$$^{***}$
& $-0.0087$$^{***}$ & $-0.0078$$^{***}$ \\
& ($0.00246$) & ($0.00069$)
& ($0.00077$) & ($0.00074$) \\[2mm]

Response text ($X_{pu}$)               & \xmark      & \xmark       & \xmark      & \xmark      \\[1mm]
Instrument ($Z_{pu}$)                  & \xmark      & \xmark       & \xmark      & \cmark      \\[1mm]
Opinion fixed-effects ($\tau_p$)       & \xmark      & \xmark       & \cmark      & \cmark      \\[1mm]
No. of debates                         & $1,144,478$ & $1,144,478$  & $1,137,968$ & $1,137,968$ \\[1mm]
$R^2$                                  & $0.054$     & $0.013$      & $0.192$     & --- \\
\bottomrule
\end{tabular}
\vspace{3mm}

Note: Standard errors (clustered by opinion) displayed in parentheses. $^{***}p<0.001; ^{**}p<0.01; ^{*}p<0.05$
\vspace{0mm}
\end{table*}%
\addtolength{\tabcolsep}{-3mm}

\clearpage

\section*{Appendix D: Instrumental Variable Specification Reduced-Form}

\addtolength{\tabcolsep}{14.5mm}
\begin{table*}[!h]
\vspace{5mm}
\small
\centering
\caption{\textbf{Reduced-form estimates.} Mean past position as an instrument for reputation.}
\begin{tabular}[t]{lc}
\toprule
& \multicolumn{1}{c}{Dependent Variable: Reputation $Y_{pu}$}\\
\midrule
Mean past position $Z_{pu}$            & $-0.0002$ ($0.00001$)$^{***}$                  \\
Skill $s_{pu}$ (percentage)            & $\phantom{-}0.0037$ ($0.00007$)$^{***}$        \\
Position $t_{pu}$ (std. deviations)    & $-0.0097$ ($0.00086$)$^{***}$                  \\
Opinion fixed-effects ($\tau_p$)       & \cmark                                       \\
No. of debates                         & $1,019,469$                                  \\
$R^2$                                  & $0.20$                                       \\
\bottomrule
\end{tabular}
\label{tab:reducedform}
\vspace{3mm}

Note: Standard errors displayed in parentheses. $^{***}p<0.001; ^{**}p<0.01; ^{*}p<0.05$
\vspace{0mm}
\end{table*}%
\addtolength{\tabcolsep}{-14.5mm}

\end{document}